\documentclass[lettersize,journal]{IEEEtran}
\usepackage{amsmath,amsfonts}
\usepackage{algorithmic}
\usepackage{algorithm}
\usepackage{array}
\usepackage[caption=false,font=normalsize,labelfont=sf,textfont=sf]{subfig}
\usepackage{textcomp}
\usepackage{stfloats}
\usepackage{url}
\usepackage{verbatim}
\usepackage{graphicx}
\usepackage{cite}

\usepackage{graphicx}
\usepackage{array}
\usepackage{makecell}
\usepackage{lipsum}
\usepackage{multirow}
\usepackage{caption}
\usepackage{geometry}
\usepackage{booktabs}
\usepackage{amssymb}
\usepackage{adjustbox}
\usepackage{hyperref}

\renewcommand{\arraystretch}{1.2}

\usepackage{hyperref}

\usepackage[justification=centering]{caption}
\usepackage[utf8]{inputenc}

\usepackage{xcolor}

\usepackage{amsmath}
\usepackage{graphicx}
\usepackage{array}
\usepackage{booktabs}
\usepackage{multirow}

\usepackage{amssymb} 
\usepackage{pifont} 
\usepackage{multirow}
\usepackage{graphicx}

\begin{document}

\title{\huge A Comprehensive Survey of Large AI Models for Future Communications:  Foundations, Applications and Challenges}

\author{Feibo Jiang, \textit{Senior Member, IEEE}, Cunhua Pan, \textit{Senior Member, IEEE}, Li Dong, Kezhi Wang, \textit{Senior Member, IEEE}, Merouane Debbah, \textit{Fellow, IEEE}, Dusit Niyato, \textit{Fellow, IEEE}, and Zhu Han, \textit{Fellow, IEEE}

	\thanks{
		Feibo Jiang (jiangfb@hunnu.edu.cn) is with the Hunan Provincial Key Laboratory of Intelligent Computing and Language Information Processing, Hunan Normal University, Changsha, China.
		
		Cunhua Pan (cpan@seu.edu.cn) is with the National Mobile Communications Research Laboratory, Southeast University, Nanjing, China.
		
		Li Dong (Dlj2017@hunnu.edu.cn) is with Xiangjiang Laboratory, Hunan University of Technology and Business, Changsha, China.
		
		Kezhi Wang (Kezhi.Wang@brunel.ac.uk) is with the Department of Computer Science, Brunel University of London, UK.

        Merouane Debbah (merouane.debbah@ku.ac.ae) is with KU 6G Research Center. Department of Computer and Information Engineering, Khalifa University, Abu Dhabi 127788, UAE.

       Dusit Niyato (dniyato@ntu.edu.sg) is with the College of Computing and Data Science, at Nanyang Technological University, Singapore.

    Zhu Han (hanzhu22@gmail.com) is with the Department of Electrical and Computer Engineering, University of Houston, Houston, TX, USA.

    GitHub link: \url{https://github.com/jiangfeibo/ComLAM}.
       	
	}

}
\maketitle

\begin{abstract}
The 6G wireless communications aim to establish an intelligent world of ubiquitous connectivity, providing an unprecedented communication experience. Large artificial intelligence models (LAMs) are characterized by significantly larger scales (e.g., billions or trillions of parameters) compared to typical artificial intelligence (AI) models. LAMs exhibit outstanding cognitive abilities, including strong generalization capabilities for fine-tuning to downstream tasks, and emergent capabilities to handle tasks unseen during training. Therefore, LAMs efficiently provide AI services for diverse communication applications, making them crucial tools for addressing complex challenges in future wireless communication systems. This study provides a comprehensive review of the foundations, applications, and challenges of LAMs in communication. First, we introduce the current state of AI-based communication systems, emphasizing the motivation behind integrating LAMs into communications and summarizing the key contributions. We then present an overview of the essential concepts of LAMs in communication. This includes an introduction to the main architectures of LAMs, such as transformer, diffusion models, and mamba. We also explore the classification of LAMs, including large language models (LLMs), large vision models (LVMs), large multimodal models (LMMs), and world models, and examine their potential applications in communication. Additionally, we cover the training methods and evaluation techniques for LAMs in communication systems. Lastly, we introduce optimization strategies such as chain of thought (CoT), retrieval augmented generation (RAG), and agentic systems. Following this, we discuss the research advancements of LAMs across various communication scenarios, including physical layer design, resource allocation and optimization, network design and management, edge intelligence, semantic communication, agentic systems, and emerging applications. Finally, we analyze the challenges in the current research and provide insights into potential future research directions.

\end{abstract}


\section{Introduction}
	
\begin{figure}[htpb]
	\centering
	\includegraphics[width=0.5\textwidth,height=0.11\textwidth]{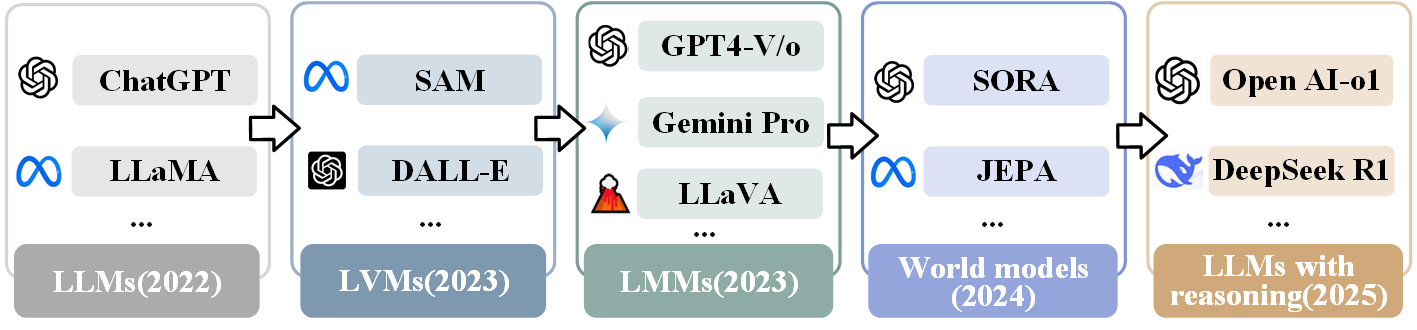}
	\caption{\textcolor{black}{The development history of LAMs.}}
	\label{fig:fig01}
\end{figure}

With the continuous emergence of various new technologies, the complexity and diversity of communication systems are steadily increasing, leading to a growing demand for efficiency, stability, and intelligence in these systems. Ubiquitous intelligence is one of the key visions for 6G, which aims to provide real-time artificial intelligence (AI) services for both the network and its users, enabling on-demand AI functionality anytime and anywhere. To achieve this, the 6G network architecture must consider the deep integration of information, communication and data technologies, building a comprehensive resource management framework that spans the entire life cycle of computing, data, AI models, and communication\cite{9606720}. \textcolor{black}{Currently, AI technology has advanced from the era of deep learning to large AI models (LAMs), such as large language models (LLMs), large vision models (LVMs), large multimodal models (LMMs) and world models}. The development history of LAMs is Shown in Fig. \ref{fig:fig01}. These LAMs possess powerful cognitive abilities, enabling efficient AI services for differentiated communication application scenarios, becoming potent tools for addressing complex challenges in future wireless communication systems. Against this backdrop, the application of LAMs in communications has become a research hotspot. This paper aims to provide a comprehensive review of the fundamentals, applications, and challenges related to LAMs for communications. 

\subsection{Background}
 6G aims to create an intelligent and interconnected world, offering an unprecedented communication experience to human society. In international mobile telecommunications for 2030 (IMT 2030), proposed by ITU-R, six typical scenarios are defined: immersive communication, ultra-massive connectivity, ultra-reliable and low-latency communication, integrated sensing and communication, ubiquitous connectivity across terrestrial, aerial, and satellite networks, and integrated AI and communication\textcolor{black}{\cite{9349624}}. Communication, sensing, computing, AI, security, and other multidimensional elements will be integrated into 6G to provide users with more advanced communication services \cite{8766143}.

To achieve the aforementioned vision, 6G relies on a range of novel communication technologies, including intelligent reflecting surfaces \cite{9847080}, integrated terrestrial and non-terrestrial networks \cite{9275613}, terahertz communications \cite{6005345}, integrated sensing and communication \cite{9737357}, digital twins \cite{9899718}, the metaverse \cite{9880528}, and quantum communication technologies \cite{cozzolino2019high}. However, the development of these new technologies has posed challenges for communication systems, such as performance approaching theoretical limits and difficulties in adapting to large-scale, complex scenario changes \cite{9144301}. The integration of AI with communication will be an effective way to address these issues. Currently, classical methods such as traditional machine learning, deep supervised learning, and deep reinforcement learning (DRL) have already been widely applied in 5G as effective tools for optimizing traditional algorithms and operations, being extensively used in core networks, transport networks, radio access networks, and edge networks. Below, we first review the development history of integrated AI and communication.

\subsubsection{Deep learning-assisted communication}

\textcolor{black}{The rapid development of deep learning has provided a solid foundation for tackling critical challenges in wireless communications \cite{8666641,yu2024multi,xu2021ordinary,zhou2021rcnet}. 
By applying deep learning techniques, communication systems have reached a new level in terms of performance and efficiency. These advancements not only improve operational capabilities but also pave the way for future innovations in communication technologies. However, in dynamic and uncertain environments, the generalization ability of deep learning is limited, and communication systems still face challenges related to adaptive optimization and learning\cite{8666641}.}

\subsubsection{Reinforcement learning-assisted communication}

\textcolor{black}{Reinforcement learning has been effectively utilized to enable communication network entities to derive optimal policies, including decisions or actions, in given states \cite{8714026,shui2023cell,wang2023ddpg,huang2023performance}. 
Therefore, reinforcement learning-based communication technologies demonstrate tremendous potential in addressing critical issues such as policy optimization, efficiency enhancement, and performance improvement in dynamic environments, thus laying a solid foundation for the continuous optimization and adaptive learning of communication systems\cite{8714026}.}

\subsubsection{Generative AI-assisted communication} 

\textcolor{black}{With the continuous advancement of AI technologies, particularly represented by transformer models, human society is rapidly entering a new era of generative AI (GAI). The development of GAI has also brought new opportunities to communications\textcolor{black}{\cite{karapantelakis2024survey}}. These generative models, including generative adversarial networks (GANs), transformers, and diffusion models, can more accurately learn the intrinsic distribution of information and possess stronger generation and decision-making capabilities, thereby significantly enhancing the performance and efficiency of communication systems \cite{10623395,zhou2023conditional,grassucci2023generative,9632815}. 
However, as communication systems become increasingly complex and the communication environment undergoes dynamic changes, GAI may encounter challenges such as mode collapse and catastrophic forgetting in high-dimensional and complex data generation tasks \textcolor{black}{\cite{manduchi2024challenges}}.}

\subsection{Motivation}
	
\subsubsection{Definition}
\begin{figure}[htpb]
	\centering
	\includegraphics[width=0.3\textwidth,height=0.3\textwidth]{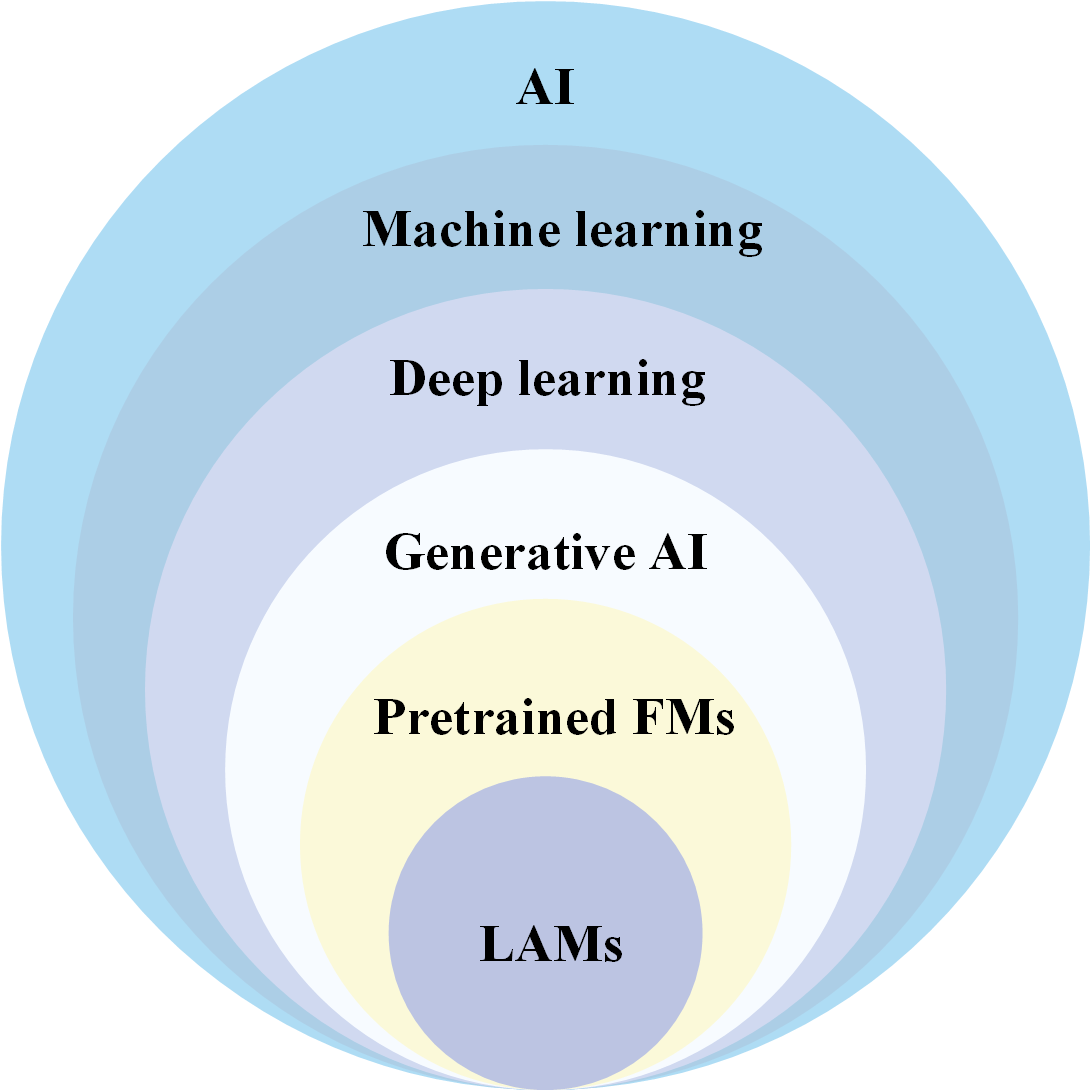}
	\caption{\textcolor{black}{The role of LAMs in AI.}}
	\label{fig:fig02}
\end{figure}

\textcolor{black}{LAMs represent cutting-edge advancements in AI, characterized by state-of-the-art generative architectures and parameter scales reaching hundreds of billions or even trillions. These models exhibit cognitive capabilities comparable to humans, enabling them to handle increasingly complex and diverse data generation tasks. Based on the modality of data they process, LAMs encompass LLMs, LVMs, LMMs, and world models\cite{wang2025large}. In recent years, several prominent LAMs such as GPT\cite{achiam2023gpt}, Sora \cite{liu2024sora}, \textcolor{black}{large language model Meta AI (LLaMA)} \cite{touvron2023llama}, and Gemini \cite{team2023gemini} have transformed workflows across various domains, including natural language processing (NLP) and computer vision.} The role of LAMs in AI is shown in Fig. \ref{fig:fig02}.

\subsubsection{Distinction between LAMs and GAIs}
\textcolor{black}{Compared to other GAI models, LAMs offer significant advantages in scale and capability. While GAI models also focus on producing new data, LAMs are typically much larger, with parameter counts reaching hundreds of billions or even trillions, and exhibit superior generalization abilities. They demonstrate greater adaptability and flexibility across a broader range of tasks. Moreover, unlike smaller GAI models, LAMs are capable of exhibiting emerging behaviors such as in-context learning\cite{10447527}, chain of thought (CoT) \cite{wei2022chain}, reflection\cite{wang2022self}, and emergence\cite{10384606}. These capabilities enable them to rapidly adapt to various downstream applications without the need for task-specific retraining.}

\subsubsection{Distinction between LAMs and pretrained FMs}
\textcolor{black}{Pretrained foundation models (FMs) have undergone extensive pre-training but have not been adapted to specific tasks. As a result, they are prone to generating hallucinations and typically require further fine-tuning to evolve into task-effective LAMs \cite{zou2024telecomgpt}. For instance, language-based FMs often need additional processes such as instruction fine-tuning and reinforcement learning from human feedback (RLHF) to develop into fully functional LLMs \cite{dong2024rlhf}. Therefore, compared to generic pre-trained FMs, LAMs can be further optimized on domain-specific datasets—such as those in communication—effectively mitigating hallucination issues inherent to raw pre-trained FMs and enabling more efficient handling of diverse communication tasks.}

Consequently, the integration of LAMs with communications offers the following distinct advantages \cite{jiang2024large2}:

\begin{itemize}
    \item {\textbf{\textcolor{black}{Exceptional global insight and decision-making:}} Future communication systems will operate in dynamic environments, influenced by device mobility and traffic fluctuations. Traditional AI methods, reliant on localized features, are prone to local optima and struggle to learn long-term spatiotemporal characteristics. LAMs, with advanced architectures and hundreds of billions of parameters, capture network features from a global perspective, adapt to multi-scale spatiotemporal dependencies, and generate stable decision-making responses, whereas traditional neural networks require retraining. For example, LAMs learn user mobility and traffic fluctuations from global perspectives, mitigating long-term forgetting and enabling precise traffic prediction and resource allocation \cite{zhang2024generative}.

    \item {\textbf{\textcolor{black}{Strong robustness and generalization:} }} Future communication systems are expected to support a diverse range of devices, such as internet of things (IoT) devices and UAVs, while offering management strategies including beamforming design, user association, and edge resource allocation. Traditional AI approaches, which focus on learning task-specific features, are constrained in their adaptability and robustness across multiple tasks. LAMs, trained on a variety of data and tasks, exhibit enhanced generalization capabilities for multitask scenarios, enabling effective decision-making in novel use cases. The extensive data allows LAMs to capture complex patterns and subtle distinctions in heterogeneous devices and imbalanced datasets. For example, by learning channel state information (CSI) and network topology, LAMs can design universal offloading models in mobile edge computing systems, optimizing task offloading and resource scheduling through prompts without requiring retraining \cite{dong2023lambo}.

 \item {\textbf{\textcolor{black}{Advanced understanding and emergent abilities:}}} Future communication systems are required to deliver tailored solutions for diverse application scenarios. For instance, autonomous driving demands ultra-low latency and high reliability, while IoT necessitates support for massive connectivity. Traditional AI approaches, reliant on small-scale models trained for specific contexts, exhibit limited applicability. Leveraging their superior contextual learning capabilities, LAMs can proactively analyze user demands and preferences in 6G networks, comprehending various scenarios with minimal or even zero-shot samples, thereby providing personalized services. Their emergent capabilities enable LAMs to perform advanced cognitive tasks such as logical reasoning and causal inference, dynamically planning, configuring, and optimizing communication networks \cite{jiang2024large3}.
}
\end{itemize}

\begin{table*}[t]
\small
\centering
\caption{\textcolor{black}{Comparison of our work with existing studies}}
\label{tab:1}
\resizebox{\textwidth}{!}{
\renewcommand{\arraystretch}{1.2}
\begin{tabular}{|m{0.5cm}<{\centering}|m{0.5cm}<{\centering}|m{1.1cm}<{\centering}|m{1.0cm}<{\centering}|m{1.2cm}<{\centering}|m{1.4cm}<{\centering}|m{1.5cm}<{\centering}|m{1.5cm}<{\centering}|m{1.5cm}<{\centering}|m{1.2cm}<{\centering}|m{1.1cm}<{\centering}|m{6.8cm}<{\raggedright}|}
\hline
Ref. & Year & Type & Model training (C1) & Evaluation metric (C2) & \makecell{Model\\architecture\\(C3)} & \makecell{Model\\classification\\(C4)} & \makecell{Model\\optimization\\(C5)} & \makecell{Application\\scenarios\\(C6)} & \makecell{Research\\challenges\\(C7)} & \makecell{Future\\directions\\(C8)} & \multicolumn{1}{c|}{Remarks} \\ 
\hline

\cite{chen2024big} & 2024 & Magazine & \checkmark &  &  &  &  & \checkmark & \checkmark & \checkmark & 
-For C2, C3, C4 and C5, the descriptions are simplistic and lack comprehensiveness. \\ 
\hline

\cite{huang2024large} & 2024 & Magazine & \checkmark &  &  &  & \checkmark &  & \checkmark & \checkmark & 
-For C2, evaluation relies on manual scoring and lacks task-specific metrics. \newline
-For C3 and C4, the focus on LLMs omits broader architectural and classification discussions. \newline
-For C6, application scenarios are insufficiently covered.\\ 
\hline

\cite{bariah2024large} & 2024 & Magazine & & & &  & \checkmark & \checkmark & \checkmark & \checkmark & 
-For C1 and C2, the work lacks coverage of emerging training methods and clear evaluation criteria. \newline
-For C3 and C4, architectural and classification content remains superficial.\\ 
\hline

\cite{zhou2024large} & 2024 & Survey &  & \checkmark & \checkmark &  & \checkmark & \checkmark & \checkmark & \checkmark & 
-For C1, only standard pre-training and fine-tuning are considered. \newline
-For C5, optimization is mentioned briefly without methodological depth.\\ 
\hline

\cite{liang2024generative} & 2025 & Survey & \checkmark & \checkmark &  & \checkmark & \checkmark &  & \checkmark & \checkmark & 
-For C3, the focus on GAI excludes newer architectures such as Mamba and SSM. \newline
-For C6, application coverage is narrow and limited to semantic scenarios.\\ 
\hline

\cite{xu2024unleashing} & 2024 & Survey & \checkmark & \checkmark &  & \checkmark &  & \checkmark & \checkmark & \checkmark & 
-For C3, traditional generative models are emphasized while recent architectures (e.g., Diffusion) are omitted. \newline
-For C5, optimization challenges are mentioned without proposing concrete solutions.\\ 
\hline

\cite{boateng2024survey} & 2024 & Survey & \checkmark &  & \checkmark & \checkmark  &  &\checkmark &\checkmark &\checkmark & 
-For C2, evaluation frameworks are lacking. \newline
-For C5, the discussion on model optimization remains relatively high-level, with limited focus on recent trends.\\ 
\hline

\cite{qu2025mobile} & 2025 & Survey &\checkmark & & &\checkmark & &\checkmark &\checkmark & \checkmark & 
-For C2 and C3, evaluation methodology and architectural coverage are insufficient. \newline
-For C5, no optimization strategies are proposed.\\ 
\hline

\cite{celik2024dawn} & 2024 & Survey & \checkmark &\checkmark &  &  &   & \checkmark & \checkmark & \checkmark & 
-For C3, architectural categorization is unclear. \newline
-For C4 and C5, classification and optimization discussions are limited in depth and integration.\\ 
\hline

\multicolumn{2}{|c|}{\textbf{Our Work}} & Survey & \checkmark & \checkmark & \checkmark & \checkmark & \checkmark & \checkmark & \checkmark & \checkmark & 
-For C1 to C8, it provides a comprehensive and recent review. It includes in-depth discussion, analyses, challenges, and future research directions.\\ 
\hline
\end{tabular}
}
\end{table*}

\subsection{Related survey work}

\textcolor{black}{Table} \ref{tab:1} compares this study with the existing related survey researches. 
\textcolor{black}{Existing surveys typically focused only on the basic principles of LAMs and some key technologies, with limited analysis of the structures and characteristics of different types of LAMs. Moreover, the coverage of the latest applications is often insufficient, particularly in the review of other LAMs beyond LLMs in communication. Although these studies have made valuable contributions to exploring the application of LLMs and GAIs for communications, there is still a need for further improvement. The limitations of existing survey studies can be summarized as follows:}
\subsubsection{Limited model coverage} \textcolor{black}{Most existing surveys primarily focus on LLMs (e.g., GPT and LLaMA), while paying insufficient attention to other categories of LAMs such as LVMs (e.g., SAM and DINO), LMMs (e.g., \textcolor{black}{Composable Diffusion (CoDi)} and ImageBind), and world models (e.g., Sora and JEPA). These studies often lack a unified framework for understanding the diverse architectures, training paradigms, and alignment strategies across different types of LAMs, resulting in an incomplete view of the model landscape in communication.}
\subsubsection{Incomplete application landscape} \textcolor{black}{While prior surveys have provided valuable insights into specific applications of LLMs in communication, their coverage of broader application scenarios remains limited. In particular, the roles and potential of other types of LAMs across diverse communication tasks (e.g., physical layer design, resource allocation, network management, edge intelligence, semantic communication, and agentic systems) have not been fully explored. Furthermore, systematic comparisons of different models’ suitability, technical characteristics, and collaboration strategies in these scenarios are largely absent, which may hinder the development of a comprehensive understanding of LAMs in communication.}

\subsection{Contributions}	

\begin{figure*}[htpb]
	\centering
	\includegraphics[width=0.7\textwidth,height=1.2\textwidth]{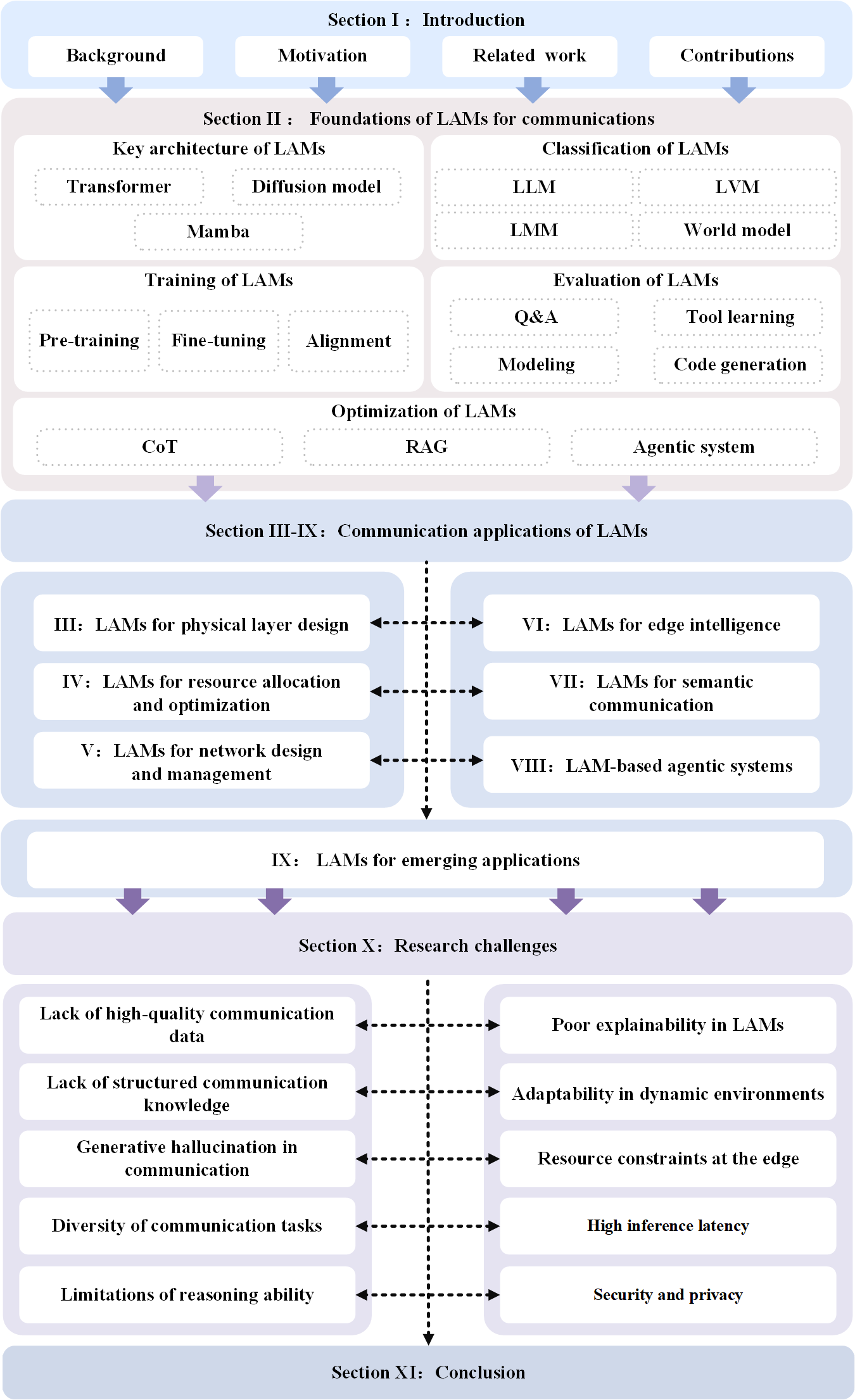}
	\caption{\textcolor{black}{Overall organization of the survey.}}
	\label{fig:fig1}
\end{figure*}

Through a comprehensive summary and systematic analysis of the existing literature, this work provides readers with a complete knowledge framework of LAMs for communications, covering the fundamental review, application review, and challenges and future directions. \textcolor{black}{Fig. \ref{fig:fig1} presents the
organization of this paper.} Specifically, the contributions of this paper are summarized as follows:

\subsubsection{Foundations of LAMs for communications}

 First, we introduce the key architectures of LAMs, including the transformer model, diffusion model, and mamba model. Next, we classify LAMs in detail, covering categories such as LLMs, LVMs, LMMs, and world models. Then, we deeply explore the pre-training, fine-tuning and alignment methods of LAMs for communications. Next, the evaluation methods are introduced, including communication question-and-answer (Q\&A) evaluation, communication tool learning evaluation, communication modeling evaluation, and communication code design evaluation. Finally, we introduce the optimization techniques of LAMs, including CoT, retrieval augmented generation (RAG), and agentic systems. These technologies can further improve the performance of the LAMs and make it effectively applied in communication. Please refer to \textcolor{black}{Section }\ref{sec:foundations} for details.

\subsubsection{Applications of LAMs for communications}

We provide a detailed overview of the research progress of LAMs in various application scenarios, \textcolor{black}{including physical layer design, resource allocation and optimization, network design and management, edge intelligence, semantic communication, agentic systems, and emerging applications.} We classify and summarize the research progress in each direction, and we detail the representative works that integrate LAMs with communications, showcasing the current research status and application prospects, and providing researchers with referenceable research directions. Detailed contents can be found in \textcolor{black}{Sections} \ref{sec:physical}-\ref{sec:emerging}.

\subsubsection{Research challenges of LAMs for communications}

\textcolor{black}{We analyze the major research challenges faced by LAMs in communication-oriented scenarios. First, the current communication landscape lacks high-quality labeled data, and issues related to privacy and cost further constrain data availability, thereby hindering model training and generalization. Second, LAMs struggle to incorporate structured domain knowledge in communications, which limits their performance in tasks such as channel modeling. Issues such as generative hallucinations, insufficient reasoning capabilities, and poor interpretability further undermine their reliability and transparency in critical tasks. Moreover, LAMs still face limitations in adaptability and generalization when dealing with dynamic network environments and diverse communication tasks. Deployment in resource-constrained scenarios remains challenging due to high inference latency, as well as concerns regarding privacy and security risks.} 
Detailed contents can be found in \textcolor{black}{Section} \ref{sec:challenges and future}.

\begin{table*}[t]
	\centering
	\small
	\caption{\textcolor{black}{List of common abbreviations}}
	\resizebox{\textwidth}{!}{
		\begin{tabular}{|c|c|c|c|}
			\hline
			\textbf{Abbr.} & \textbf{Description} & \textbf{Abbr.} & \textbf{Description} \\ \hline
			AGI & Artificial General Intelligence & BitFit & Bias-Only Fine-Tuning \\ \hline
			CoDi & Composable Diffusion & CoT & Chain of Thought \\ \hline
			Diffusion & Diffusion Model & DPO & Direct Preference Optimization \\ \hline
			FM & Foundation Model & GAN & Generative Adversarial Network \\ \hline
			GPT & Generative Pretrained Transformer & ICL & In-context Learning \\ \hline
			KD & Knowledge Distillation & LLM & Large Language Model \\ \hline
			LLaMA & Large Language Model Meta AI & LMM & Large Multimodal Model \\ \hline
			LLaVA & Large Language and Vision Assistant & LoRA & Low-Rank Adaptation \\ \hline
			LVM & Large Vision Model & MAS & Multi-Agent Systems \\ \hline
			MoE & Mixture of Experts & MPFT & Mixed-Precision Fine-Tuning \\ \hline
			PEFT & Parameter-Efficient Fine-Tuning & PPO & Proximal Policy Optimization \\ \hline
			PTQ & Post-Training Quantization & RAG & Retrieval-Augmented Generation \\ \hline
			RLHF & Reinforcement Learning from Human Feedback & RLAIF & Reinforcement Learning from AI Feedback \\ \hline
			SFT & Supervised Fine-Tuning & SLM & Small Language Model \\ \hline
			SSM & State Space Model & VLM & Vision-Language Model \\ \hline
			WM & World Model & WS & World Simulator \\ \hline
		\end{tabular}
	}
\end{table*}

\section{Foundations of LAMs for communications}
\label{sec:foundations}

\textcolor{black}{Compared to traditional AI and machine learning models, LAMs are built with hundreds of billions or even trillions of parameters and advanced architectures. Through large-scale pretraining, they achieve strong multi-task generalization. LAMs exhibit superior cognitive abilities and multimodal reasoning capabilities, enabling rapid adaptation to downstream tasks via in-context learning and fine-tuning. Moreover, their emergent abilities allow them to understand and solve tasks that are not explicitly seen during training. This generality and adaptability give LAMs a distinct advantage in the development of future intelligent communication systems. Then, we present the foundations of LAMs for communications, including key architectures, model classifications, model training and evaluation, and optimization methods. }
\subsection{Key architecture of LAMs}

 LAMs demonstrate exceptional capabilities in handling complex data and tasks through continuous optimization and innovation. The key architectures play a crucial role in the successful application of LAMs, significantly enhancing the performance and efficiency of the LAMs while also fostering ongoing advancements in related technologies. This section introduces the key architectures of LAMs and their research advancements in communication, including the transformer model, diffusion model, and mamba model.

\subsubsection{Transformer model} 
Transformer is a novel neural network architecture introduced by Vaswani et al. in 2017\cite{vaswani2017attention}. The primary characteristic of the transformer architecture lies in its complete reliance on the attention mechanism, which eliminates the sequential dependencies inherent in traditional sequence data, allowing the model to process input sequences in parallel. This architecture excels in addressing long-range dependency issues, particularly in NLP tasks. Compared to traditional recurrent RNNs, the transformer exhibits superior parallelism and computational efficiency, making it well-suited for handling large-scale datasets and complex sequence tasks. The workflow of the transformer architecture is as follows:

        \begin{itemize}	
		\item {\textbf{Input embedding and positional encoding:}} The transformer converts each word in the input sequence into high-dimensional vectors through an embedding layer, which represents their semantic information. Positional encodings are added to these word vectors to enable the model to recognize the order of the sequence and perceive the sequential relationships.
        
		\item {\textbf{Multi-head self-attention in the encoder:}} The processed word vectors enter the multi-head self-attention layer of the encoder. The self-attention layer computes attention weights through queries, keys, and values, determining the relevance of each word to others and capturing global dependency information in the sequence. The output is then further processed through a feedforward neural network, with residual connections and layer normalization applied to enhance the stability and efficiency of model training.
       
		\item {\textbf{Generation and output in the decoder:}} The hidden state vectors generated by the encoder are passed to the decoder, which first processes the previously generated portions of the output sequence to capture the internal dependencies of the current sequence. Subsequently, the cross-attention mechanism integrates the current state of the decoder with the hidden states from the encoder, generating new outputs based on the input sequence. Finally, after processing through the output layer, the decoder produces the final output sequence.
        
	\end{itemize}

\textcolor{black}{\emph{Representative transformer-based research in communication:}} The transformer has been widely adopted in various cutting-edge LAMs, such as OpenAI's GPT series, Google's BERT and T5 models, and Facebook's RoBERTa\cite{liu2019roberta}. In recent years, transformer models have been increasingly utilized in communication due to their exceptional global modeling capabilities and efficient parallel computing performance. For instance, Wang et al. \cite{wang2022transformer} explored the application of the transformer architecture in handling large-scale MIMO and semantic communication in 6G networks, highlighting the critical role of deep learning, particularly the transformer, in network optimization and the resolution of intricate communication challenges. Yoo et al. \cite{yoo2022real} proposed a real-time semantic communication system based on a vision transformer (ViT), which significantly outperformed traditional 256-QAM systems in low signal-to-noise ratio environments, demonstrating the advantages of semantic communication in effectively conveying information. Furthermore, Wu et al. \cite{wu2024transformer} introduced the DeepJSCC-ViT-f framework, which leverages the ViT architecture in conjunction with channel feedback information to enhance the performance of wireless image transmission. This framework aims to address the complexities and adaptability issues of the existing joint source-channel coding (JSCC) methods.

\subsubsection{Diffusion model} The Diffusion model is a generative model based on the probabilistic diffusion process, proposed by Sohl-Dickstein et al. in 2015 \cite{sohl2015deep}. The main feature of the diffusion model is to generate data by gradually adding noise to the data and then learning the inverse denoising process. It is good at generating high-quality, detailed images, especially when dealing with complex image generation and signal recovery problems. The workflow of the diffusion model is as follows:

\begin{itemize} 
 \item {\textbf{Forward diffusion process:}} The forward diffusion process maps the data to a state close to the standard normal distribution by gradually adding gaussian noise. This process gradually destroys the data, making the data more and more blurred, and finally forming a high-noise state. The noise addition at each step is gradual, gradually covering up the original structural information of the data.
       
 \item {\textbf{Reverse diffusion process:}} In the reverse diffusion process, the model restores the data from the high-noise state to the state of the original data by gradually removing the noise. This process is usually approximated by training a neural network to the probability distribution of the reverse diffusion process. The network learns how to gradually recover the data from the noise to generate new samples similar to the original data. This step is the key to the generation process, enabling the model to effectively ``reconstruct" the data from the noise.
         
\end{itemize}

\emph{Representative diffusion model-based research in communication:} In recent years, diffusion models have drawn widespread attention due to their exceptional performance and flexibility. These models have demonstrated particularly outstanding performance in image generation and show considerable application potential in communication. \textcolor{black}{For instance, Jiang et al.\cite{jiang2024large1} proposed a channel estimation method called generative adversarial network and diffusion model aided-channel estimation (GDCE), combining a diffusion model and a CGAN. CGAN is first used to generate CSI. Then a diffusion model refines the CSI information. Gradually removing noise through the diffusion model generates a more refined CSI image, improving signal recovery accuracy. Du et al.\cite{du2023beyond} studied the applications of generative diffusion models (GDMs) in network optimization, especially in complex intelligent network scenarios like DRL, incentive mechanism design, semantic communication, and vehicular networks, demonstrating GDMs' potential for modeling complex data distributions and generating high-quality decisions. Wu et al.\cite{wu2023cddm} proposed channel denoising diffusion models (CDDM) for image transmission tasks in wireless semantic communication systems. This model enhances image transmission quality and communication system performance by leveraging diffusion models' noise-removal advantages through learning channel input signals' distribution. Additionally, Duan et al.\cite{duan2024dm} introduced a diffusion model for multiple-input multiple-output (DM-MIMO) module for robust MIMO semantic communication. It combines diffusion models with SVD precoding and equalization to reduce channel noise, lower MSE, and enhance image reconstruction quality, showing superior performance in MIMO systems.}  Chi et al. \cite{chi2024rf} proposed RF-Diffusion, a radio signal generation model based on time-frequency diffusion theory. Through an enhanced diffusion process, this model is capable of generating high-quality sequential RF signals in both the time and frequency domains. \textcolor{black}{Grassucci et al. \cite{grassucci2024diffusion} proposed a generative audio semantic communication framework using a conditional diffusion model to address bandwidth consumption and error recovery challenges in traditional audio signal transmission.}

\subsubsection{Mamba model} Mamba is a generative architecture for efficiently processing long sequence data, proposed by Gu et al. in \textcolor{black}{2022} \cite{gu2023mamba}. The main feature of mamba is to efficiently process long sequence data. It allows the model to focus on relevant information and filter out unnecessary parts through a selection mechanism based on input data. At the same time, it adopts a hardware-aware computing algorithm, which specifically optimizes the processing performance on the GPU and significantly speeds up computing. The mamba model is good at processing high-dimensional, long-series complex data, such as natural language, video, or time series tasks. By optimizing data flow processing and resource allocation, it can effectively reduce communication delays and improve system performance. The workflow of the mamba architecture is given as follows:

\begin{itemize}
    \item \textbf{Input processing and projection:} Input data (such as text, images, time series, etc.) is segmented into multiple fragments (tokens or patches) and converted to vector representations through a linear projection layer. This step is similar to the preprocessing process of other deep learning models and is used to map the input into a high-dimensional space.

    \item \textbf{Selection mechanism:} The state space is a collection of variables that describe the dynamic behavior of the model. Mamba uses an efficient selection mechanism to dynamically adjust the state space parameters based on the input data. This mechanism allows the model to filter out irrelevant information and retain only key feature information, thereby achieving content-aware modeling. \textcolor{black}{This process is usually implemented using a convolutional layer}

    \item \textbf{SSM calculation:} 
    State space model (SSM) calculation is the process of modeling input data and generating output using the SSM. The discretized SSM equation is used to calculate the input data, which includes the state equation and observation equation. The state equation describes the change of state variables over time, and the observation equation describes how the observation variable is generated from the state variable. Mamba architecture uses these equations to learn complex patterns in sequence data and generate high-quality output.

    \item \textbf{Output generation:} After the SSM completes processing the input, mamba passes the output to a fully connected layer or other task-related layers (such as a classifier or generator) to generate the final output.

\end{itemize}

\emph{Representative mamba-based research in communication: }Mamba architecture has made significant breakthroughs in long sequence modeling and multi-task parallel computing. Its efficient processing capabilities and dynamic adjustment mechanism have received widespread attention. \textcolor{black}{For instance, Wu et al\cite{wu2024mambajscc} proposed the MambaJSCC architecture for wireless transmission of images. The architecture implements adaptive coding based on the generalized state space model (GSSM) and CSI residual transfer (CSI-ReST). MambaJSCC uses the VSSM-CA module and combines the reversible matrix transformation of GSSM to effectively capture global information and achieve high performance and low latency in image transmission without increasing calculation and parameter overhead, surpassing existing methods. Yuan et al\cite{yuan2024st} proposed ST-Mamba, which addresses the accuracy and stability issues of traffic flow estimation, and performs well, especially when data is limited. ST-Mamba combines the CNN and the mamba framework, which can effectively capture the spatial and temporal characteristics of traffic flow. Yu et al \cite{yu2024vimsc} proposed the underwater acoustic image semantic communication model VimSC based on vision mamba. By combining OFDM technology, the quality of image transmission in complex underwater environments is significantly improved. VimSC uses a CSI feedback precoding network to adjust semantic symbols and uses channel estimation and equalizers to achieve accurate image reconstruction. Experimental results show that its performance is better than traditional methods in low signal-to-noise ratio environments.}

\subsection{Classification of LAMs}

\textcolor{black}{As shown in Table \ref{tab:Classification}, we classify LAMs into the following categories based on the type of data processed \cite{carolan2024review}. Although other classification methods have been proposed in previous studies, our data-type-based classification provides a more focused and practical framework, which is better suited to addressing the diverse challenges in communication systems, such as handling different modalities, optimizing resource allocation, and improving system efficiency across various communication tasks.}
\subsubsection{Large language model} LLM is an NLP model with a large number of parameters and complex architecture. It learns the structure and semantics of language by pre-training on a large amount of text data. These models can generate natural and fluent text and perform a variety of language tasks such as translation and Q\&A. LLMs are usually based on deep learning architectures, such as transformers, which can effectively capture long-range dependencies. They adjust internal parameters and improve performance by optimizing complex loss functions. The LLM includes the following technical features:

\begin{itemize}
\item \textbf{Language understanding and generation:} LLMs show powerful language understanding and generation capabilities when processing text data. By pre-training on large-scale texts, they learn rich language patterns and knowledge, and can understand complex language structures and contexts. LLMs not only identify and interpret the meaning of words, phrases, and sentences but also capture nuances in language, such as tone and emotion. When generating text, they create coherent and creative content, maintain grammatical and semantic accuracy, and are capable of multilingual translation, demonstrating the potential for cross-language understanding\textcolor{black}{\cite{wu2025survey}}. 
            
\item \textbf{Memory and reasoning ability:} LLMs are widely used because of their excellent memory and reasoning capabilities. Through deep learning of massive text data, they can memorize and understand rich language knowledge and factual information, and can maintain consistency and coherence in different contexts. The model not only masters vocabulary and grammar, but also understands complex context and long-distance dependencies. In terms of reasoning, LLMs can perform logical reasoning based on text, infer implicit meanings, causal relationships and conclusions, handle multi-step reasoning tasks, and simulate the human thinking process to a certain extent. They use information in memory to reason and predict new situations, generating coherent and logical text, making them excellent at tasks such as summary generation, Q\&A and text analysis.

        \end{itemize}
   
   Classic LLMs include the GPT series, the Gemma series, and the LLaMA series, among others. These models possess a vast number of parameters, enabling them to process and generate natural language text effectively while demonstrating excellent performance across various NLP tasks. Below, we provide a detailed overview of three classic LLMs. 
  
  \begin{itemize}
\item \textbf{GPT series:} The GPT series is developed by OpenAI, standing for ``generative pretrained transformer". These models learn language patterns and generate natural language through pretraining on a vast amount of text data. Since the introduction of GPT-1, the GPT models have evolved through several versions, including GPT-1 \cite{radford2018improving}, GPT-2 \cite{radford2019language}, GPT-3 \cite{brown2020language}, GPT-4 \cite{achiam2023gpt}, and OpenAI o1. The original GPT-1, released in 2018, focused on text generation, and learning language patterns from extensive text data using unsupervised learning. In 2019, GPT-2 was released, expanding the parameter count from 100 million to 1.5 billion, resulting in more coherent text generation and the ability to handle more complex tasks. Subsequently, GPT-3 was launched in 2020, with a further parameter increase to 175 billion, showcasing powerful few-shot learning capabilities, enabling it to perform various tasks such as translation, question answering, and code writing without fine-tuning. The release of GPT-4 in 2023 introduced multimodal capabilities, allowing it to understand images in addition to text, along with significant improvements in reasoning ability, logic, and coherence, making it adept at tackling complex reasoning problems. 
\textcolor{black}{In 2024, OpenAI released model o1, which demonstrates exceptional capabilities in reflective reasoning compared to previous generations of LAMs. It is able to generate more precise and logically consistent responses by conducting multi-level analyses of complex problems. This allows o1 to effectively perform self-correction and reflection in ambiguous or uncertain situations, thereby enhancing its reliability and intelligence in practical applications.}
The advancements in the GPT series have opened new possibilities for the development of NLP and AI.

\item\textbf{Gemma series:} The Gemma series is developed by Google, which includes Gemma 1 \cite{team2024gemma} and Gemma 2 \cite{team2024gemma2}. Gemma 1 was released in \textcolor{black}{2024} and is available in two versions with different scales: 2 billion parameters and 7 billion parameters, catering to various computational environments and application needs. The model architecture is based on the transformer decoder and incorporates several technical improvements, such as multi-head attention mechanisms, rotary position embedding (RoPE), and GeGLU activation functions, enabling the model to exhibit strong contextual understanding and excel in diverse text generation tasks. Gemma 2 was launched in 2024, offering versions with 9 billion and 27 billion parameters. This model features an enhanced transformer architecture, which includes the interleaved use of local and global attention, as well as group query attention techniques, thereby improving the model's language processing capabilities. Compared to Gemma 1, Gemma 2 demonstrates significant advancements in both parameter scale and performance.

 \item  \textbf{LLaMA series:} The LLaMA series is a foundational language model developed by the Meta AI team, which includes the LLaMA-1 \cite{touvron2023llama}, LLaMA-2 \cite{touvron2023llama2}, and LLaMA-3 \cite{dubey2024llama}. Both LLaMA-1 and LLaMA-2 were released in 2023. The LLaMA-1 series features several models of varying sizes, ranging from 700 million to 6.5 billion parameters. Compared to previous language models, LLaMA-1 enhances the accuracy and fluency of text generation while maintaining low computational costs through optimization of the model structure and the introduction of improved algorithms during training. LLaMA-2 builds on this foundation with significant improvements, expanding the model's scale to offer versions with 700 million, 1.3 billion, 3 billion, and 7 billion parameters. It enhances the model's contextual understanding by optimizing the self-attention mechanism and algorithms. LLaMA-3 was released in 2024, further extending the parameter count to offer versions with 8 billion and 70 billion parameters, as well as incorporating additional pre-training data, demonstrating superior performance in benchmark tests.

        \end{itemize}

\emph{LLM-based research in communication: } LLMs, due to their powerful data processing capabilities, have been widely applied in communication, enhancing the efficiency of communication systems and significantly promoting their rapid development\cite{yang2025wirelessgpt}. The integration of LLMs with communication systems is also a crucial research direction for the future.
\textcolor{black}{For example, Jiang et al.\cite{jiang2024large5} proposed a LAM-driven multimodal semantic communication (LAM-MSC) framework. It can achieve multimodal to unimodal data conversion, personalized semantic extraction and wireless CSI estimation, thus addressing data discreteness and semantic ambiguity issues in multimodal semantic communication by LLMs.}
\textcolor{black}{ In addition,Jiang et al.\cite{jiang2024semantic} proposed a semantic communication architecture based on FMs. They introduced GPT, leveraging LLMs to enhance semantic understanding and data reconstruction efficiency, and employed fine - tuning techniques to address high computational complexity from numerous model parameters.
Jiang et al.\cite{jiang2024large3} proposed a multi-agent system to enhance LLMs' capabilities in 6G, improving network optimization and management via natural language input. It integrates data retrieval, collaborative planning, and evaluation to address limited private communication data and constrained reasoning, thereby extending LLMs' knowledge and functionality in this context. \textcolor{black}{Wang et al.\cite{10570717} presented a general end-to-end learning semantic communication model integrating LLMs to boost next-generation communication systems' performance. It addresses challenges like semantic fidelity, cross-scenario generalization and complexity via subword-level tokenization, rate adapters for channel codec matching and fine-tuning for private background knowledge. Xu et al.\cite{xu2024large2} proposed a split learning system for LLMs agents in 6G networks to enhance human machine interaction and provide personalized cross-domain assistant services. The system offloads complex tasks to edge servers to address mobile device capacity limits. Its architecture with perception, grounding, and alignment modules enables inter module communication to meet diverse 6G user requirements.} }

\begin{table*}[t]
    \small
    \centering
    \caption{Classification of LAMs and their applications in communication}
    \label{tab:Classification}
    \renewcommand{\arraystretch}{1.5}
    \setlength{\tabcolsep}{10pt}
    \begin{tabular}{|>{\centering\arraybackslash}m{3cm}|>{\centering\arraybackslash}m{2cm}|>{\centering\arraybackslash}m{4cm}|>{\centering\arraybackslash}m{6cm}|}
        \hline
        \textbf{LAM Category} & \textbf{Data Type Processed} & \textbf{Specific Models} & \textbf{Application Domains} \\ 
        \hline
        Large Language Model & Text data & GPT series, Gemma series, LLaMA series & Semantic communication\cite{jiang2024large5}, Network management\cite{jiang2024large3}, Edge intelligence\cite{xu2024large2}, Security and privacy\cite{jiang2024large3} \\ 
        \hline
        Large Vision Model & Image data & SAM series, DINO series, Stable diffusion series & Semantic communication\cite{jiang2024large2}, Image segmentation\cite{zhang2024segment}, Image generation\cite{zhang2024vision} \\ 
        \hline
        Large Multimodal Model & Multimodal data & CoDi series, Meta-transformer, ImageBind & Semantic communication\cite{jiang2024large5}, Cross-Modal Generation\cite{zhang2023meta} \\ 
        \hline
        World Model / World Simulator & Simulated real-world data & Sora, Vista, JEPA & Autonomous driving\cite{gao2024vista}, Digital twin\cite{liu2024sora}\\ 
        \hline
    \end{tabular}
\end{table*}

\subsubsection{Large vision model}
LVM is a foundation model that processes and understands visual data. It usually adopts CNN and transformer architecture. The LVM learns rich visual features from a large number of images and can demonstrate high accuracy and strong generalization capabilities in tasks such as image classification, object detection, image segmentation, and generation. With continuous development and optimization, LVMs play an important role in promoting the advancement of image processing technology\textcolor{black}{\cite{zhang2024vision}}.
The LVM has many technical features, such as feature representation learning and support for multiple visual tasks as follows:
 \begin{itemize}
     \item \textbf{Feature representation learning:} In LVMs, feature representation learning is one of the core technologies that automatically extracts and learns important features in images through deep neural networks. This process mainly relies on CNN and ViT to complete. The CNN first extracts local features through multi-layer convolution and nonlinear activation functions, and then integrates these local features into global features through fully connected layers or pooling operations. The transformer architecture further enhances the capability of feature representation through a self-attention mechanism, capturing long-range dependencies and complex contextual information in images. Through large-scale pre-training and careful fine-tuning, the LVM can optimize feature representation, and significantly improve the performance of visual tasks. 
    
\item \textbf{Support multiple visual tasks:} Supporting multiple visual tasks is one of the important features of LVMs. Through deep learning technology, it can support multiple visual tasks and play a role in a wide range of application scenarios. These tasks include image recognition, object detection, scene parsing, image segmentation, image generation, image editing, and video analysis. They can recognize and understand objects and scenes in images, locate the position and size of objects, analyze the relationship between objects, segment image areas, create or modify image content, and process actions and events in videos. In addition, they also support 3D reconstruction to enhance the visual experience of virtual environments.

 \end{itemize}
   
    Classic LVMs include the SAM series, DINO series, Stable Diffusion series, etc. In the following, we provide a detailed introduction to these LVMs.
   
 \begin{itemize}
     \item \textbf{SAM series:} SAM is an LVM developed by Meta AI, designed to perform image segmentation efficiently. The SAM series includes SAM-1\cite{kirillov2023segment} and SAM-2\cite{zhang2024segment}. SAM-1 was released in 2023. Its core technology is a deep learning architecture based on the self-attention mechanism, which can recognize any object in the image and refine its boundaries with high resolution. The model is designed with a wide range of application scenarios, and can not only handle conventional target segmentation tasks, but also complex multi-target segmentation and detail processing. SAM-2 was released in 2024 and has been improved in many aspects to further improve the performance of image segmentation. First, SAM-2 optimizes segmentation accuracy, especially when dealing with complex scenes and small targets, and can more accurately recognize and segment multiple types of objects. Secondly, SAM-2 has been upgraded in model architecture, introducing more advanced deep learning algorithms and optimized self-attention mechanisms, enabling it to more effectively capture details and long-range dependencies in images. In addition, the inference speed has also been improved, and the processing efficiency is higher, especially in scenarios that require real-time response.
    
\item \textbf{DINO series:} DINO Series is an unsupervised visual feature learning model jointly developed by Meta AI Research and Inria. It is designed to generate universal visual features through large-scale curated data sets without the need for fine-tuning. This series of models include DINO V1\cite{caron2021emerging} and DINO V2\cite{oquab2023dinov2}. DINO V1 was released in 2021. It uses the transformer architecture and adopts a contrastive learning method. By inputting images from different perspectives for processing, DINO V1 can learn to recognize and distinguish different elements and structures in the image. This approach allows DINO V1 to be pre-trained on unlabeled image data and generate powerful image representations suitable for various vision tasks, such as image classification, object detection, etc. DINO V2 was released in 2023. Compared with DINO V1, DINO V2 made significant improvements in many aspects. DINO V2 adopts a more advanced architecture, expands the model size, and uses more computing resources, thereby improving the accuracy of feature representation and the ability to handle complex visual tasks. The contrastive learning strategy and self-supervision mechanism are optimized to improve the robustness and generalization ability for different image types. During the training process, DINO V2 introduced improved training techniques and data enhancement methods to enhance its performance in complex scenes and small target processing.

\item \textbf{Stable diffusion series:} The stable diffusion series is developed by Stability AI for generating high-quality images. These models use diffusion model technology and are widely used in tasks such as image generation, image restoration, and image transformation. This series include Stable Diffusion V1\cite{rombach2022high}, Stable Diffusion V2\cite{rombach2022high}, and Stable Diffusion V3\cite{esser2024scaling }. Stable Diffusion V1, released in 2022, is able to generate delicate and diverse images through a large amount of training data and diffusion model technology. This model marks an important breakthrough in the field of image generation, with the ability to generate high-resolution images in a variety of scenarios. Then, Stable Diffusion V2 was released in \textcolor{black}{2022}, bringing even more significant improvements. This version uses updated generation technology to support higher resolution images and perform better at handling complex scenes and details. Stable Diffusion V3 was released in 2024. Compared with V2, Stable Diffusion V3 replaces the U-Net backbone of V2 by introducing a rectified fourier transformer (RFT) architecture, which significantly improves image and text encoding processing capabilities. Stable Diffusion V3 uses three chip encoding tracks (i.e., original text encoding, converted text encoding, and image encoding) to improve multimodal interaction with images, enabling the generation of more refined and contextually accurate images, especially for complex cues. 

 \end{itemize}
    
\emph{LVM-based research in communication: }    By applying LVMs to the communication field, communication systems can be made more efficient in processing visual tasks. 
\textcolor{black}{Jiang et al.\cite{jiang2024large2} utilized the SAM to construct a semantic knowledge base (SKB), thereby proposing the LAM-SC framework, a semantic communication framework based on LAMs, focused on the transmission of image data. The SAM performs accurate semantic segmentation on any image without specific training, breaking the image down into multiple segments, each containing a single semantic object. Additionally, Tariq et al.\cite{tariq2023segment} proposed a SAM-based semantic communication method to retain semantic features and ensure high-quality reconstruction in image transmission. This method leverages the SAM to overcome the diminishing returns of traditional approaches aimed at enhancing transmission rates while reducing communication overhead. }

\subsubsection{Large multimodal model} LMMs are capable of simultaneously processing and understanding data from different modalities, such as vision, language, haptic and auditory. These models achieve comprehensive processing and reasoning of multimodal information by integrating the features of various modalities in a unified high-dimensional space. They utilize advanced neural network architectures, such as transformers and diffusion models, to extract features from each modality and optimize their representations through techniques \textcolor{black}{such as} contrastive learning and self-supervised learning. By training across multiple modalities, these models are able to understand and relate the semantic relationships between different modalities, thereby demonstrating superior performance in handling complex multimodal data and providing intelligent, efficient solutions. Unlike visual language models (VLMs), LMMs support modalities beyond vision and text\cite{wu2023multimodal}.

LMMs show strong capabilities in processing multimodal information. Their core technical features are cross-modal fusion and multimodal representation learning:

 \begin{itemize}
     \item \textbf{Multimodal representation learning:} Multimodal representation learning is an important technology of LMMs. It integrates feature representations of different modalities such as images, speech and text into a unified high-dimensional space. First, the LMM uses ViT to extract image features and uses a transformer to extract text and speech features. Then, these high-dimensional vectors are fused through methods such as splicing and weighted summation to form a unified feature representation. This fusion enables the LMM to better understand and associate information from different modalities, improving the performance of multimodal tasks.

     \item \textbf{Cross-modal fusion:} LMMs integrate multiple data types such as text, images, audio and video through cross-modal fusion technology to achieve deeper understanding and analysis. These LMMs can process data from different modalities at the same time and learn the relationship between them. For example, the LMM can combine images with related text to generate richer descriptions; in video analysis, it can understand the visual content as well as the voice and text information in the video. In addition, these LMMs can also perform cross-modal reasoning and prediction, such as generating images or audio from text. Such capabilities make LMMs widely used in NLP, computer vision, speech recognition and other fields.

\end{itemize}
   
LMMs integrate many advanced model architectures and can process and understand data in different modalities. In the following, we provide a detailed introduction to three LMMs.

\begin{itemize}
    \item \textbf{CoDi series: }CoDi series is developed by Microsoft Azure and the University of North Carolina. It is an innovative multi-modal generative model. The series include CoDi-1\cite{tang2023any}, CoDi-2 \cite{tang2024codi}. CoDi-1 was launched by Microsoft in 2023, aiming to improve the accuracy and flexibility of image generation. CoDi-1 utilizes conditional diffusion model technology to achieve precise control of the generated results by combining specific condition information (such as text descriptions, labels, or other input data) with the image generation process. CoDi-2 was released in 2024. Compared with CoDi-1, CoDi-2 has made significant improvements in many aspects, further improving the ability and effect of image generation. First, CoDi-2 introduces an enhanced conditional control mechanism, allowing the generated images to more accurately conform to complex conditional inputs. This improvement includes a more flexible condition encoding method and a more refined condition processing strategy, thus providing higher control accuracy. Secondly, CoDi-2 optimizes the model architecture by adopting more advanced diffusion technology, and improves network design, making the generated images of higher quality and richer in detail. In addition, CoDi-2 introduces improved data augmentation methods and optimized training techniques, resulting in enhancements in the speed and stability of image generation.

    \item \textbf{Meta-transformer:} Meta-transformer\cite{zhang2023meta} is a multimodal learning framework designed to process and associate information from different modalities. It uses a fixed encoder to achieve multimodal perception without paired multimodal data. The framework consists of three main components: a unified data segmenter that maps data of various modalities to a shared latent space; a modality-shared encoder that extracts high-level semantic features; task-specific heads. Meta-transformer can uniformly process 12 modalities, such as natural language, images, point clouds, audio, video, infrared, hyperspectral, X-ray, time series, tables, inertial measurement units (IMUs), and graphic data. Its main advantage is that it converts data of different modalities into a unified feature sequence, uses a shared encoder to extract features, reduces the complexity of cross-modal alignment, and improves the flexibility of training.

    \item \textbf{ImageBind:} ImageBind\cite{girdhar2023imagebind} is an advanced LMMs that aims to integrate data from different modalities through a shared embedding space. The model can handle data from six different modalities, such as images, text, audio, depth, thermal imaging, and IMU data. Its innovation lies in the cross-modal alignment without explicit pairing of data. Through contrastive learning, data from different modalities are projected into a unified representation space, thereby enhancing the generalization ability and cross-modal understanding ability of the model. ImageBind performs well in multimodal retrieval, classification, and generation tasks, especially when dealing with unaligned data.

\end{itemize}
 
\emph{LMM-based research in communication: } LMMs are widely used in communication due to their powerful multimodal information processing capabilities\cite{jiang2025commgpt}. For example, Jiang et al.\cite{jiang2024large5} proposed a LAM-MSC framework by combining the multimodal processing model CoDi with the language communication system. In this communication system framework, the CoDi model can convert multimodal data into text data for processing to achieve cross-modal processing of the model. The LAM-MSC framework shows excellent performance in simulation experiments, and can effectively process multimodal data communication and maintain the semantic consistency between the original data and the restored data. Qiao et al.\cite{qiao2024latency} proposed a delay-aware semantic communication framework based on a pre-trained generative model by combining models such as BLIP, Oscar, and GPT-4. The framework aims to achieve ultra-low data rate semantic communication in future wireless networks through multimodal semantic decomposition and transmission. In this framework, the transmitter performs multimodal semantic decomposition on the input signal and selects appropriate encoding and communication schemes to transmit each semantic stream according to the intention. For text prompts, a retransmission-based scheme is adopted to ensure reliable transmission, while other semantic modalities use adaptive modulation/coding schemes to adapt to the changing wireless channels.

\subsubsection{World model} \textcolor{black}{A world model is an abstract framework to describe and simulate real-world phenomena, aiming to create intelligent systems that can understand and simulate the environment\textcolor{black}{\cite{10655190}}. The world model primarily consists of two key components: the environment simulator and the controller. The environment simulator is responsible for constructing a model that can predict the state and behavior of the environment, typically achieved through deep neural networks. These networks are trained to understand the dynamic characteristics of the environment and generate predictions of future states and rewards\textcolor{black}{\cite{9729537}}. The controller uses this simulator to make decisions and improves its performance in the real environment by training and optimizing in the simulated environment.}

\textcolor{black}{World models support LAMs by providing simulated scenarios that help LAMs generalize and adapt to complex and dynamic environments. Unlike digital twins, which are primarily used to replicate real-world objects or systems in real time, world models focus on simulating and training LAMs in virtual environments\textcolor{black}{\cite{10538211}}. We introduce the features of the world model in details below.}

\begin{itemize}
    \item \textbf{Long-term planning and cognitive decision-making:} The world model simulates and predicts the dynamic changes of complex systems and makes effective decisions. Long-term planning involves learning patterns from historical data and anticipating future trends to guide resource allocation and action selection. World models can evaluate the long-term impact of different strategies and help decision-makers understand different choices and formulate sustainable plans. The world model can also simulate the decision-making process in different scenarios, provide a variety of solutions, and support wise choices in complex environments. This dynamic and predictive ability makes it valuable in policy formulation, resource management, and risk assessment.

    \item \textbf{Continuous perception and embodied intelligence:} The world model has significant advantages in continuous perception and embodied intelligence. It can obtain information from the environment in real time, and monitor and analyze various variables, such as climate, traffic flow, etc., to provide the latest data for decision-making. Embodied intelligence enables models to combine sensory information with physical entities to simulate the behavior and interaction of entities in the environment. This capability supports more complex tasks such as automatic control, robot navigation, and environmental monitoring, giving it broad application prospects in areas such as intelligent transportation, smart city management, and disaster warning.

\end{itemize}   

   There are many classic world models, which provide many new ideas for the research in communication. In the following, we provide a detailed introduction to three world models.

 \begin{itemize}
     \item \textbf{Sora:} Sora is a groundbreaking text-to-video generation model released by OpenAI\cite{liu2024sora} that demonstrates significant emergent capabilities. It is based on a pre-trained diffusion transformer and is able to generate high-quality videos based on text instructions, introducing details through progressive denoising and text cues. Sora excels in several areas, including simulation capabilities, creativity, and accessibility. Although not explicitly 3D modeled, Sora demonstrates 3D consistency, such as dynamic camera motion and long-range coherence, and is able to simulate aspects of the physical world and simple interactions. 
     
\item \textbf{JEPA:} Joint embedding predictive architecture (JEPA)\cite{lecun2022path} is a world model for multi-modal learning that aims to enhance the understanding of complex data through joint embedding and prediction tasks. By mapping different modalities of data into a shared embedding space, JEPA enables the model to capture the potential relationships between different data in this space. Specifically, JEPA performs contrastive learning in the embedding space and optimizes the embedding distance of similar data to enhance the understanding of different modal information. In addition, in the interaction between JEPA and the environment, the world model can provide generated samples and state changes, and JEPA further adjusts the structure and characteristics of its embedded space through this dynamic information, allowing it to reason more effectively in complex environments. This interactive mechanism not only improves the understanding of the environment, but also enhances the adaptability of JEPA, allowing it to exhibit higher robustness and flexibility in diverse real-world scenarios.

\item \textbf{Vista:} Vista\cite{gao2024vista} is an advanced world model focused on solving the limitations of data scale, frame rate, and resolution in the field of autonomous driving. It adopts a novel loss function to enhance the learning of moving instances and structural information, and designs a latent replacement method to achieve coherent long-term predictions through historical frames. Vista also excels when it integrates a diverse set of controls from high-level intentions to low-level actions. After large-scale training, Vista outperforms most existing video generation models in experiments on multiple datasets. Vista's training framework includes two stages: high-fidelity future prediction and multi-modal action control learning, which can provide high-resolution predictions in different scenes and camera angles with less quality degradation.

 \end{itemize}
   
  \emph{World model-based research in communication: }  The application of world models in communication has played a revolutionary role in 6G. For example, Saad et al.\cite{saad2024artificial} proposed a revolutionary vision of the next-generation wireless system in their research, the AGI-native wireless system, the core of which is the world model. The AGI-native wireless system is mainly built on three basic components: the perception module, the world model, and the action planning component. Together, these components form four pillars of common sense, including handling unforeseen scenarios through horizontal generalization, capturing intuitive physics, performing analogical reasoning, and filling in the gaps. The study also discussed how AGI-native networks can be further utilized to support three use cases related to human users and autonomous agent applications: analogical reasoning for next-generation digital twins, synchronous and elastic experience of cognitive avatars, and brain-level metaverse experience with holographic transmission as an example. Finally, they put forward a series of suggestions to inspire the pursuit of AGI-native systems and provide a roadmap for next-generation wireless systems beyond 6G.
    
\subsection{Training of LAMs for communications}
The training process of LAMs for communications involves three stages: pre-training, fine-tuning, and alignment. As illustrated in Table \ref{tab:learning}, a comprehensive comparison of these stages is provided. In the following sections, we present a detailed discussion of each stage.
\subsubsection{Pre-training of LAMs for communications}The pre-training stage forms the foundation for LAMs to acquire specialized knowledge in communication.  This process is summarized as follows:

The LAMs is pre-trained on a large unlabeled dataset to learn universal features, boosting performance on communication tasks, reducing reliance on labeled data, and improving knowledge transfer. The key pre-training methods are self-supervised and multi-task learning:
\begin{itemize}	
\item \textbf{Self-supervised learning:}
Self-supervised learning, unlike unsupervised learning, enables the LAMs to learn features from the data itself by generating supervisory signals through data transformation or masking. The process starts with data preprocessing, followed by creating proxy tasks to generate self-supervised signals. The model is then trained using these internal representations, similar to supervised learning but without external labels \cite{liu2024llm4cp}. 

\item \textbf{Multi-task learning:}
Multi-task learning improves model performance by learning multiple related tasks simultaneously. Tasks share model parameters, enabling the LAMs to leverage their relationships for better efficiency and generalization. The process involves defining tasks, designing a shared model architecture with common and task-specific layers, and ensuring consistent data preprocessing. During training, shared layers capture common features, while task-specific layers focus on individual objectives \cite{mohammed2023deep}. 
\end{itemize}

 To improve training efficiency and model performance, researchers have proposed various optimization strategies for the pre-training stage:
\begin{itemize}
\item \textbf{Distributed training: }Distributed training techniques involve multiple devices working together to train LAMs, requiring effective data and model parallelism strategies for efficiency and stability. Frameworks like megatron-lm and deepspeed are designed for distributed training, enabling efficient data and model parallelism \cite{smith2022using}.

\item \textbf{Learning rate scheduling:}
Dynamic learning rate adjustment plays a crucial role in enabling LAMs to identify optimal parameters during training. Typical approaches include cosine annealing and cyclic learning rate strategies \cite{li2020cyclical}.

\item \textbf{Gradient clipping:}
This optimization technique mitigates gradient explosion and vanishing by scaling or truncating gradients during backpropagation. Typical approaches include absolute value clipping and norm-based clipping, which constrain or reduce excessively large gradients \cite{zhang2019gradient}.

\end{itemize}
\subsubsection{Fine-tuning of LAMs for communications}
The fine-tuning stage optimizes a pre-trained LAM using a specific communication dataset, helping it better adapt to communication tasks. This process improves the model's understanding, generalization, accuracy, and efficiency in communication applications.

 Telecom instruction fine-tuning technique\cite{zou2024telecomgpt} trains LLMs to generate accurate outputs based on telecom instructions in natural language. It uses instructions paired with responses to guide the model in performing tasks, enhancing its understanding and ability to handle new tasks. The instruction dataset is generated using advanced LLMs like GPT-4 and LLaMA-3, based on telecom documents, to meet the needs of various tasks \cite{zou2024telecomgpt}:
\begin{itemize}
    \item \textbf{Multiple choice question answering:} Selecting all correct answers from a set of multiple-choice questions.
    \item \textbf{Open-ended question answering:} Providing open-ended responses to telecom-related questions based on standards, research papers, or patents.
    \item \textbf{Technical document classification:} Classifying the text of various technical documents into relevant working groups, such as the different working groups in the 3GPP standards.
    \item \textbf{Mathematical modeling:} Generating accurate mathematical equations, such as channel models, based on textual descriptions of system models and problem statements.
    \item \textbf{Code generation:} Generating scripts or functions for specific tasks or functionalities in telecom.
    \item \textbf{Fill-in-the-middle:} Completing incomplete scripts based on context and target functionality.
    \item \textbf{Code summarization:} Summarizing the core functionalities of a given script, including identifying whether the script is related to telecom.
    \item \textbf{Code analysis:} Detailing the operational logic behind functions, emphasizing knowledge and principles relevant to telecom.
   
\end{itemize}

Based on the designed instruction fine-tuning dataset, the steps of fine-tuning LAMs for communications are as follows:

\begin{itemize}
\item \textbf{Model initialization:} After creating the instruction fine-tuning dataset, select a pre-trained LAM as the initial model, ensuring it has strong language understanding and generation capabilities for communications.

\item \textbf{Model adjustment and optimization:} 
Use the instruction-response pairs dataset for supervised fine-tuning (SFT) of the pre-trained LAM, learning the relationship between instructions and responses while adjusting model parameters. Then, define a negative log-likelihood loss function to measure the difference between the model's generated responses and the expected ones \cite{zou2024telecomgpt}.

\item \textbf{Iterative training:}
Through multiple iterations, the LAM learns to generate quality responses based on instructions. It updates its parameters using the loss function after processing each batch of instruction-response pairs. 

\item \textbf{Final evaluation and application:}
After training, the LAM is evaluated to ensure it meets performance standards across tasks. It is then tested in real-world scenarios for practicality and reliability before being deployed in communication applications.
\end{itemize}

There are various techniques for fine-tuning LAMs, including LoRA, Adapters, BitFit, and Prefix Tuning as follows.
\begin{itemize}
    \item \textbf{LoRA} \cite{hu2021lora} (Low-rank adaptation) is an efficient fine-tuning method that reduces computational and storage costs while maintaining model performance. By limiting weight matrix updates to a low-rank subspace, it decreases the number of parameters updated, improving fine-tuning efficiency without compromising task performance. 

    \item \textbf{Adapter}\cite{houlsby2019parameter} is a fine-tuning method that adds small, trainable modules at each layer of the LAM, keeping the pre-trained model parameters fixed. This reduces the number of parameters to update, saves resources, and supports multi-task learning, making it ideal for resource-limited scenarios.

    \item \textbf{BitFit}\cite{zaken2021bitfit} (Bias-only fine-tuning) significantly reduces computational and storage costs by updating only the bias terms in the LAM. It minimizes parameter updates, maintains performance, and adapts quickly to specific tasks, without requiring complex changes to pre-trained models.

    \item \textbf{Prefix tuning}\cite{li2021prefix} fine-tunes pre-trained LAMs by adding a trainable prefix vector to the input sequence, while keeping the model's original weights fixed. It reduces computational and storage costs by updating only the prefix, making it efficient for adapting to specific tasks.
\end{itemize}

The fine-tuning stage helps LAM better understand and execute communication instructions without explicit examples, improving its ability to respond accurately to communication tasks and enhancing its effectiveness in real-world applications.
\subsubsection{Alignment of LAMs for communications}
Alignment tuning is a crucial step to better align the LAM's responses with human preferences. After SFT on a communication dataset, the LAM may still generate undesirable responses, such as repetition, overly short replies, or irrelevant content. Key alignment techniques can address these issues.

Alignment tuning improves model performance by guiding the LAM to generate more accurate and reasonable responses. RLHF \cite{dong2024rlhf} is a form of alignment fine-tuning that combines human feedback with traditional reinforcement learning to optimize LAM performance. RLHF is especially useful in communication tasks, where decision-making and output reliability are critical, enabling more efficient learning of complex tasks. The RLHF workflow typically involves several key steps.:
\begin{itemize}	
\item \textbf{Environment and agent construction:} Develop a fundamental reinforcement learning framework that includes both the environment (alignment task) and the agent (LAM).

\item \textbf{Human feedback collection:} Collect feedback from human experts during the agent's task execution through interactive methods, including performance evaluations, suggestions, or corrections.

\item \textbf{Reward modeling:} Convert human feedback into reward signals and train a reward model using machine learning to accurately interpret and quantify the feedback into appropriate reward values.

\item \textbf{Reinforcement training:} Use reward signals from the reward model to train the agent with reinforcement learning, updating its strategy to gradually optimize performance and better align with human expectations.

\end{itemize}

In addition to RLHF, there are also key alignment technologies such as reinforcement learning from AI feedback (RLAIF), proximal policy optimization (PPO), and direct preference optimization (DPO) as follows. 
\begin{itemize}
    \item \textbf{RLAIF} \cite{lee2023rlaif} is a new method for improving the behavior of LAMs. Unlike traditional RLHF, RLAIF uses AI-generated feedback to optimize models, reducing the need for large human-annotated datasets. AI agents (e.g., GPT-4) evaluate model outputs and adjust parameters based on these evaluations to improve performance. The process involves two steps: first, the AI agent generates feedback by evaluating the model’s outputs, and second, this feedback is used to adjust the model through reinforcement learning, gradually improving output quality. RLAIF is more efficient and scalable, eliminating the need for costly human data.

     \item \textbf{PPO} \cite{schulman2017proximal} is a reinforcement learning method that aims to stabilize policy updates during optimization. Unlike traditional policy-gradient methods, which involve complex calculations and constraints to prevent large policy shifts, PPO uses a "surrogate objective function" and limits the update step size. PPO introduces a penalty term to control the magnitude of policy changes, ensuring that the updated policy remains close to the original. This approach improves policy performance, avoids expensive constrained optimization, and results in better convergence and robustness.

     \item \textbf{DPO} \cite{rafailov2024direct} is a reinforcement learning technique that directly optimizes model outputs to match user or system preferences, without using a reward model. By incorporating explicit preference feedback during training, DPO avoids the complexity of traditional methods and improves model performance. It is particularly effective in tasks requiring fine control of model behavior and efficient handling of complex preferences.
\end{itemize}

\begin{table*}[t]
    \centering
    \caption{Comparison of the three-stage learning process of LAMs\cite{jiang2024personalized}}
    \label{tab:learning}
    \small
    \begin{tabular}{|>{\centering\arraybackslash}m{2.6cm}|*{6}{>{\centering\arraybackslash}m{2cm}|}}
        \hline
        & \textbf{Objective} & \textbf{Data Requirement} & \textbf{Learning Approach} & \textbf{Adjusting Parameters} & \textbf{Privacy} & \textbf{Resource Requirements} \\ 
        \hline
        
        \textbf{Pre-training} 
        & Learning general language representation & Abundant and diverse datasets & Unsupervised learning & All parameters & Requires public dataset & High computational and storage resources \\ 
        \hline

        \textbf{Fine-tuning} 
        & Understanding user instructions & Instruction data & Supervised learning & Few parameters & May require user instructions & Low computational and storage resources \\ 
        \hline

        \textbf{Alignment} 
        & Aligning with user preferences & Preference-specific data & Reinforcement learning & Few parameters & May require user preferences & Low computational and storage resources \\ 
        \hline
    \end{tabular}
\end{table*}

\subsection{Evaluation of LAMs for communications}
The evaluation of LAMs for communications is a critical objective, as research on evaluation metrics not only influences the performance of LAMs but also provides deeper insights into their strengths and limitations in communication-related tasks. Selecting high-quality telecommunication datasets is essential as a prerequisite for effective evaluation. For example, Maatouk et al.\cite{maatouk2023teleqna} proposed the benchmark dataset TeleQnA to assess the knowledge of LLMs in telecommunications. The dataset consists of 10,000 questions and answers drawn from diverse sources, including telecommunication standards and research articles. Additionally, TeleQnA introduced the automated question generation framework used to create the dataset, which incorporated human input at various stages to ensure its quality.

Once an appropriate benchmark dataset has been selected, the evaluation of LAMs for communications is conducted. The evaluation framework encompasses various aspects, including communication Q\&A, communication tool learning, communication modeling, and code design.
\subsubsection{Communication Q\&A}
The evaluation of communication Q\&A \cite{zou2024telecomgpt} aims to assess the capability of LAMs, such as GPT-4, to comprehend and process communication-related documents. This task involves generating multiple-choice and open-ended questions from sources like literature, patents, and books on communication topics, including technologies, protocols, and network architectures. The performance of the LAM is measured by comparing its responses to ground-truth answers, with particular emphasis on its understanding and application of communication knowledge.

The evaluation process starts with the selection of relevant decoments, followed by data preprocessing. The LAM generates questions based on the processed content, and the generated answers are subsequently verified for accuracy, either manually or through automated comparison with standard answers. The LAM's performance is evaluated by analyzing its responses against the correct answers, focusing on accuracy as well as comprehension and reasoning abilities. Metrics such as precision, recall, and f1 score are employed to measure the quality of the answers and to assess the model's overall effectiveness in communication Q\&A tasks.

\subsubsection{Communication tool learning}
The evaluation of tool learning \cite{guo2023evaluating} examines whether LAMs can effectively select and utilize communication tools, such as existing algorithms and codes, to address real-world tasks. This capability is assessed in two key areas: tool selection, which refers to the model's ability to choose appropriate tools through reasoning, and tool usage, which involves leveraging these tools to enhance task performance, \textcolor{black}{such as integrating existing channel model code with LAMs to perform channel prediction, thereby improving the performance of communication systems.}

The evaluation emphasizes two primary aspects: the model's ability to select the correct tools and its competence in performing operations with them. This includes assessing performance with individual tools and the effectiveness of combining multiple tools, as demonstrated in benchmarks like toolalpaca \cite{tang2023toolalpaca}. These benchmarks evaluate the LAM's overall proficiency and limitations in multi-tool usage. Insights from these evaluations highlight the model's strengths and challenges in tool selection and application, guiding future optimization efforts for communication-related tasks.

\subsubsection{Communication modeling}
The evaluation of communication modeling focuses on assessing the ability of LAMs to represent and solve mathematical problems related to communication systems \cite{zou2024telecomgpt}. Tasks such as equation completion are emphasized, where critical mathematical expressions are concealed, and the physics-informed LAM must accurately predict the missing components. The evaluation begins with the selection of relevant mathematical models and equations to ensure that the tasks are both challenging and representative of real-world communication systems.

The LAM's performance is evaluated by comparing its predictions against standard answers, with particular attention to accuracy and equation consistency. Beyond precision, the evaluation also examines the model's depth of reasoning and understanding of complex communication principles. By combining measures of accuracy with an assessment of reasoning ability, this evaluation provides a comprehensive understanding of the LAM's effectiveness in tackling communication modeling tasks.

\subsubsection{Communication code design} The evaluation of communication code design \cite{zou2024telecomgpt} aims to assess the ability of LAMs to generate, complete, and analyze communication-related code in programming languages such as C, C++, Python, and Matlab. The evaluation tasks include code generation, completion, and analysis, testing the model's proficiency in creating scripts, completing partial code, and providing accurate summaries or error analyses for communication tasks.

The evaluation begins by presenting programming scenarios where the LAM (e.g., OpenAI Codex) is required to generate code for tasks such as signal processing, network protocol implementation, or data transmission algorithms. Subsequently, the LAM is tested on code completion, where it predicts and fills in missing segments, ensuring logical consistency and correct functionality. Additionally, the LAM is tasked with analyzing the given code, explaining its functionality, identifying errors, and suggesting optimizations.
Performance is measured by comparing the generated code against standard answers, with an emphasis on accuracy, completeness, and logical correctness. The model’s ability to analyze code is also evaluated, reflecting its understanding of programming concepts specific to communication.

\subsection{Optimization of LAMs}  

 To further improve the performance and adaptability of these LAMs, researchers have proposed a variety of optimization techniques, such as CoT, RAG and agent systems. In the following, we provide a detailed introduction to these optimization techniques.

\subsubsection{Chain of thought} CoT is a reasoning technique that was first proposed by Google research in 2022 \cite{wei2022chain}. The main feature of CoT is its ability to decompose complex problems into a sequence of logical reasoning steps and solve them in a linear and structured manner. It excels at solving tasks that require multi-step reasoning and comprehensive analysis, making it particularly suitable for scenarios that demand the simulation of human thought processes, such as complex decision-making and problem-solving. The workflow of the CoT method is as follows:

\begin{itemize}	

\item \textbf{Task input:} The model is presented with a complex communication task or problem, which may be provided as a natural language description, a mathematical equation, or a logical reasoning question. Based on the nature of the problem, the model identifies an appropriate reasoning pathway and integrates relevant contextual information to support the reasoning process.

\item \textbf{Logical reasoning:} The model decomposes the problem into a sequence of logical reasoning steps, performing inference in a structured, step-by-step manner. The output of each step is contingent on the results of the preceding step, ensuring a coherent and systematic reasoning process.

\item \textbf{Decision output:} The model produces a logically consistent answer or decision derived from the reasoning process. Validation mechanisms are employed to verify the correctness and reliability of the result, ensuring its accuracy and dependability.
\end{itemize}

\emph{CoT-based research in communication: }With the rapid development of the AI field, CoT technology has gradually gained widespread attention as an innovative reasoning framework. CoT helps models handle complex reasoning and decision-making tasks more efficiently by simulating hierarchical and structured reasoning in communication. \textcolor{black}{For example, Du et al.\cite{du2023power} applied CoT technology to help LLMs perform multi-step reasoning in field-programmable gate array (FPGA) development and solve complex tasks such as the implementation of fast fourier transform (FFT). CoT prompts enable LLMs to gradually decompose problems and perform calculations, improving the accuracy of generated hardware description language (HDL) code. Zou et al.\cite{zou2024genainet} used CoT technology in the GAI Network (GenAINet) framework to help distributed GAI agents perform collaborative reasoning. Agents use CoT prompts to decompose complex tasks and acquire knowledge from other agents, thereby improving decision-making efficiency and reducing communication resource consumption. Shao et al.\cite{shao2024wirelessllm} used CoT technology in the wireless LLM (WirelessLLM) framework to improve the reasoning ability of LLMs, helping the model gradually handle complex tasks in wireless communications, such as power allocation and spectrum sensing. This approach effectively enhances the task performance capabilities of LLMs in multimodal data environments.}

\subsubsection{Retrieval-augmented generation} RAG is a technology that integrates retrieval and generation, proposed by Facebook in 2020 \cite{lewis2020retrieval}. RAG combines two steps of retrieval and generation to enhance the answering ability of the LAM by retrieving relevant documents. RAG can use the retrieval module to obtain the latest and most relevant information while maintaining the powerful language capabilities of the LAM, thereby improving the accuracy and relevance of the answer. It excels at tasks that are information-rich but require knowledge from large amounts of text, such as answering questions, generating detailed instructions, or performing complex text generation. The workflow of RAG technology is given as follows:

\begin{itemize}
\item \textbf{Information retrieval:} Retrieve documents related to the input content from an external knowledge base. By using information retrieval technology, the LAM can filter out the documents that best match the input question from the knowledge base.
\item \textbf{Information fusion:} The retrieved documents are spliced with the input question as the new input of the LAM. In the information fusion stage, the LAM processes the documents and input content through the encoder, closely combines the retrieved knowledge with the question, and enhances the model's understanding and generation capabilities of the question.
\item \textbf{Generate output:} The information-fused input is passed to the LAM, and the LAM not only relies on the original input, but also uses the retrieved document information to provide richer and more accurate answers. The generation process ensures that the answer is coherent and contextually relevant, thereby ensuring the rationality and effectiveness of the output.
\end{itemize}

\emph{RAG-based research in communication: }In communication, RAG technology has demonstrated excellent application potential. For example, Bornea et al. \cite{bornea2024telco} proposed Telco-RAG, an open-source RAG system designed for the 3GPP documents, which enhances the performance of LLMs in telecom. Tang et al.\cite{tang2024automatic} proposed an automatic RAG framework for 6G use cases, using LLMs to automatically generate network service specifications, especially in the environment of an open radio access network (ORAN). Through this method, business innovators can quickly and cost-effectively evaluate and generate the required communication specifications without having an in-depth understanding of complex 6G standards, which greatly promotes innovation and application deployment in an open 6G access environment. \textcolor{black}{Huang et al.\cite{huang2024toward} proposed a 6G network-based RAG service deployment framework, aiming to improve the quality of generation services by combining LLMs with external knowledge bases. The article explores the feasibility of extending RAG services through edge computing, and proposes technical challenges in multimodal data fusion, dynamic deployment of knowledge bases, service customization, and user interaction, providing innovative directions for RAG services in future 6G networks. Xu et al.\cite{xu2024large1} proposed LMMs as a general base model for AI-native wireless systems. The framework combines multimodal perception, causal reasoning, and RAG to handle cross-layer network tasks, and experimentally verified the effectiveness of LMMs in reducing hallucinations and improving mathematical and logical reasoning capabilities. Yilma et al.\cite{gmy2024telecomrag} introduced the telecomRAG framework, which combines RAG with LLM technologies to help telecom engineers parse complex 3GPP standard documents and generate accurate and verifiable responses. By retrieving standard documents of 3GPP release 16 and release 18, the framework provides a solution with higher accuracy and technical depth than the general LLM for the telecom.}

\subsubsection{Agentic system} The agentic system is a framework consisting of LAM-based agents that perceive their environment and collaborate to achieve specific objectives. The primary characteristics of the agentic system include autonomy, adaptability, and interactivity. It can adjust its behavior according to changes in the environment and interact with other agents or the environment to optimize decision-making and task execution. It excels at solving communication problems that require dynamic responsiveness, complex decision-making, and task optimization. By simulating the behavior of humans or biological systems, agents can efficiently accomplish tasks in dynamic and changing communication environments. The workflow of the LAM-based agentic system is as follows:

\begin{itemize}	
\item \textbf{Task understanding and planning}:
The agentic system interprets input instructions, extracts relevant context, and breaks down complex tasks into smaller, manageable subtasks. It then formulates a logical plan to execute these subtasks.

\item \textbf{Execution and adaptation}:
The agent executes the planned actions, leveraging the LAM for tasks such as generating content, solving problems, or interacting with external systems. It continuously monitors progress and adapts dynamically to environmental changes or unexpected outcomes.

\item \textbf{Validation and feedback}:
The agentic system validates the results to ensure accuracy and consistency, providing reliable outputs. Feedback from the process is integrated into the system, enabling iterative improvement and enhanced performance in future tasks.
 
\end{itemize}

\emph{Agent-based research in communication: }\textcolor{black}{By leveraging its autonomy, adaptability, and interactivity, the agentic system effectively addresses complex tasks and problems, demonstrating exceptional potential in communication. For example, Tong et al.\cite{tong2024wirelessagent} proposed the wirelessagent framework, which uses LLM as the core driver and constructs a multi-agent system through four modules: perception, memory, planning, and action. The wirelessagent framework aims to solve the increasingly complex management problems in wireless networks, especially in the upcoming 6G era, when traditional optimization methods cannot cope with complex and dynamically changing network requirements. Through the collaboration and communication between multiple agents, the framework can autonomously process multimodal data, perform complex tasks, and make adaptive decisions.  Xu et al.\cite{xu2024large2} proposed a 6G network-based LLMs agent split learning system to address the issue of low local LLMs deployment and execution efficiency due to the limited computational power of mobile devices. The system achieves collaboration between mobile devices and edge servers, with division of labor between the perception, semantic alignment, and context binding modules to complete the interaction tasks between the user and the agent. Additionally, by introducing a novel model caching algorithm, the system improves model utilization, thereby reducing the network cost of collaborative mobile and edge LLMs agents. Yang et al. \cite{10815060} proposed an agent-driven generative semantic communication (A-GSC) framework based on reinforcement learning to tackle the challenges of large data volumes and frequent updates in remote monitoring for intelligent transportation systems and digital twins in the 6G era. Unlike existing semantic communication research, which mainly focuses on semantic extraction or sampling, the A-GSC framework successfully integrates the intrinsic properties of the source information with the task's contextual information. Furthermore, the framework introduces GAI to enable the independent design of semantic encoders and decoders.}

\subsection{Summary and lessons learned}
\subsubsection{Summary}
	
\textcolor{black}{This chapter provides a comprehensive overview of the key architecture, classification, training, evaluation, and optimization of LAMs in communication. First, we introduce the key architecture of LAMs. Next, we present a more detailed classification system for LAMs in communication. Following this, we discuss the training process for communications LAMs, summarizing the complete workflow from pretraining and fine-tuning to alignment, with an in-depth explanation of each of these three techniques. We then introduce the evaluation methods for communications LAMs\textcolor{black}{\cite{10769058}}, providing a comprehensive summary of the standards and metrics used to assess LAM performance in communication. Finally, we explore various optimization techniques of LAMs for communications\textcolor{black}{\cite{10650885}}. This chapter lays a solid foundation for the application of LAMs and offers clear directions for their future development.}

\subsubsection{Lessons learned}
\textcolor{black}{Although progress has been made in the construction and optimization of LAMs in communication, several lessons can be learned. Current mainstream architectures, such as transformer\textcolor{black}{\cite{10720163}}, diffusion, and mamba, demonstrate excellent modeling and reasoning capabilities. However, they still face significant difficulties in resource-constrained environments, multimodal tasks, and real-time communication applications. These challenges include high computational complexity, slow convergence, and difficulties in training, evaluation, and deployment. Regarding optimization strategies, although methods like CoT, RAG, and agentic systems have effectively enhanced the model's reasoning and task adaptability, limitations still exist in terms of stability, consistency, and efficiency.}

\section{LAMs for physical layer design}

\label{sec:physical}

\textcolor{black}{With the continuous development of wireless communication technology, especially the demand for 6G networks, physical layer design faces increasingly complex challenges. In order to meet these challenges, LAMs
and GAI models have gradually become key tools in physical layer design.}
\begin{figure*}[htpb]
		\centering
		\includegraphics[width=\textwidth,height=0.38\textwidth]{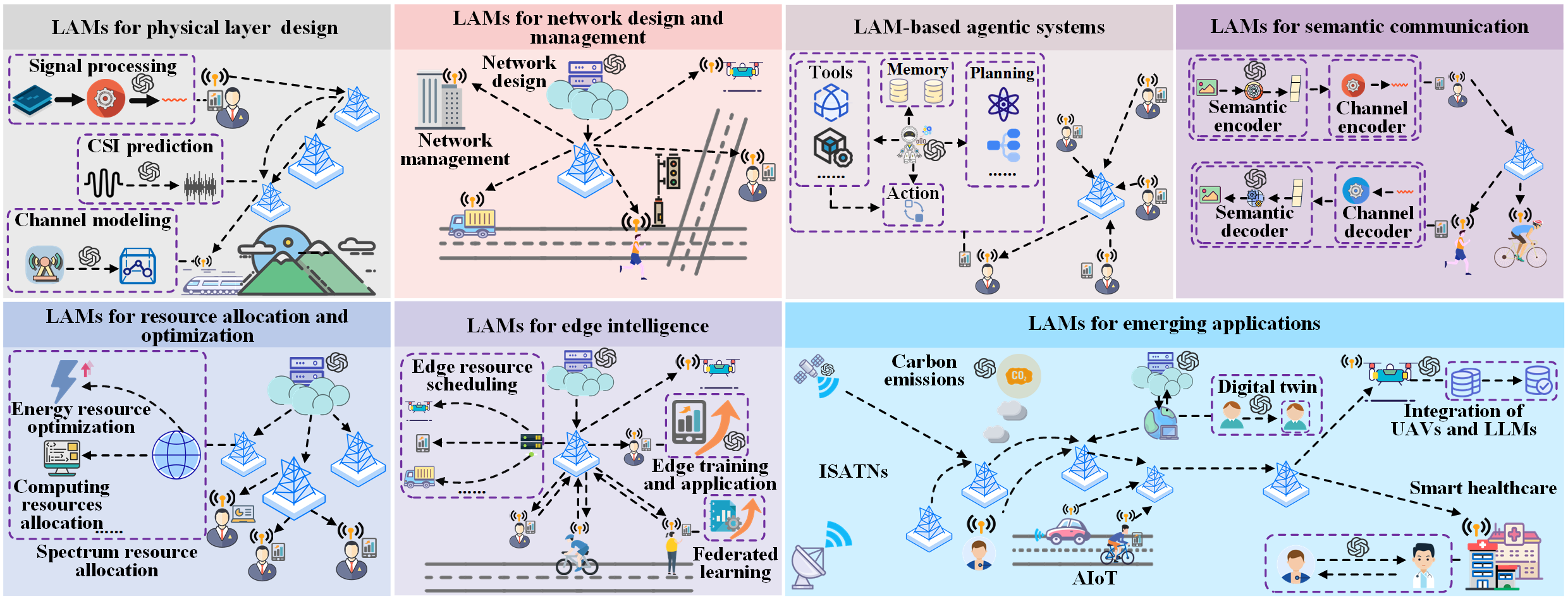}
		\caption{\textcolor{black}{Applications of LAMs in Communication. LAMs can be applied across various domains in communication, including physical layer design, resource allocation and optimization, network design and management, edge intelligence, semantic communication, agentic systems, and emerging applications.}}
           
		\label{fig:fig3}
	\end{figure*}

\subsection{Channel and beam prediction based on LAMs}
With the rapid advancement of wireless communication systems, particularly in the context of 5G and the evolution toward 6G networks, the demand for accuracy and efficiency in channel and beam prediction has grown significantly. Traditional methods often fall short when dealing with the complex and dynamic nature of modern networks. In recent years, breakthroughs in LLMs have provided new approaches to address these challenges. \textcolor{black}{For example, Fan et al. \cite{fan2024csi} proposed CSI-LLM, a method for downlink channel prediction in large-scale MIMO systems. By aligning wireless data with NLP tasks, it leverages LLMs for modeling variable-length historical sequences, showing strong performance, especially in multi-step prediction under high dynamics. Liu et al. \cite{liu2024llm4cp} proposed LLM4CP, a channel prediction method using pre-trained LLMs. It combines channel characteristic modules and cross-modal knowledge transfer for accurate TDD/FDD prediction, reducing training costs and showing strong generalization and efficiency. Sheng et al. \cite{sheng2024beam} studied beam prediction in millimeter-wave communication, using LLMs to convert time-series data into text and enhancing context with the prompt-as-prefix technique. Compared to traditional LSTM models, this approach shows stronger robustness and generalization in high-dynamic environments. Akrout et al. \cite{akrout2024multilayer} reviewed deep learning in the wireless physical layer, emphasizing trade-offs in accuracy, generalization, compression, and latency. They noted that focusing on accuracy often limits model performance in complex communication scenarios due to poor generalization.
}By analyzing the decoding tasks of end-to-end communication systems, the important impact of this trade-off on the practical application of the model is revealed, especially when LLMs are used for wireless communications, the balance between compression and latency becomes a crucial factor.

\subsection{Automated physical layer design based on LAMs}
As wireless networks continue to grow in scale and complexity, the need for intelligent and automated physical layer design has become increasingly urgent. LLMs and GAI technologies are emerging as powerful tools to address this demand, offering new possibilities for building adaptive and efficient communication systems. \textcolor{black}{For example, Xiao et al. \cite{xiao2024llm} proposed an LLM-based 6G task-oriented physical layer automation agent (6G LLM agents), introducing LLMs as intelligent co-pilots to enhance understanding and planning for dynamic tasks through multimodal perception and domain knowledge. Using a two-stage training framework, the agent effectively performs protocol question answering and physical layer task decomposition. Wang et al. \cite{wang2024generative} proposed a physical layer design framework using GAI agents that combine LLM with RAG technologies, demonstrating strong potential in signal processing and analysis. GAI agents enable rapid generation of complex channel models across environments, accelerating research in next-generation MIMO channel modeling and estimation.
}

\subsection{Summary and lessons learned}

\subsubsection{Summary}
\textcolor{black}{This chapter discusses the application of LAMs in physical layer design, demonstrating their potential in channel estimation, task decomposition, signal processing, etc. LAM can significantly improve the intelligence and automation level of the physical layer design through its powerful reasoning capabilities and multi-task learning\textcolor{black}{\cite{zheng2024large}}. The LAM improves channel estimation and blind channel equalization by accurately modeling complex data distribution. LAMs provide innovative ideas and methods for physical layer design, and are expected to bring breakthroughs in performance improvement and system optimization of future wireless communication systems\textcolor{black}{\cite{10670056}}.}

\subsubsection{Lessons learned}

\textcolor{black}{From the chapter, we have learned several important lessons. First, LAMs exhibit limited interpretability in physical layer optimization\textcolor{black}{\cite{10670056}}. Although they can generate seemingly effective optimization strategies, they often lack rigorous mathematical analysis or theoretical guarantees, which constrains their application in high-reliability communication scenarios. Second, the training and inference of LAMs rely heavily on high-quality annotated data, yet the acquisition and labeling of physical layer data are costly, making it difficult to scale data-driven LAM models in practical deployments. Therefore, enhancing interpretability and addressing data acquisition challenges are key directions for future research.}

 \section{LAMs for resource allocation and optimization}

Resource allocation and optimization are complex and critical issues in communication networks. With the development of LAMs, their application in this field has gradually shown great potential. In the following, we discuss the application of LAMs in computing resource allocation, spectrum resource allocation, and energy resource optimization.

 \subsection{Computational resource allocation} As communication networks become more complex, users have higher requirements for network services. How to provide users with high-quality communication services under limited computing resources is a major challenge. The study of computing resource allocation based on GAI models and LAM is an important research direction in the future. \textcolor{black}{ For example, Du et al. \cite{10409284} introduced the AGOD algorithm, which uses a diffusion model to generate optimal AIGC service provider (ASP) selection decisions from Gaussian noise. Combined with DRL into the D2SAC algorithm, it enhances ASP selection efficiency and optimizes user computing resource allocation. In addition, Du et al. \cite{10592370} proposed a network optimization method based on the MoE framework and LLM, leveraging LLM reasoning to manage expert selection and decision weighting, enabling efficient resource allocation and reduced energy and implementation costs. Tests on maze navigation and network service provider (NSP) utility tasks demonstrate its effectiveness in complex network optimization.}

\subsection{Spectrum resource allocation}
In current communication systems, spectrum resource allocation is an important part of achieving efficient and reliable data transmission. With the rapid development of mobile communication technology, especially in the 5G and upcoming 6G era, the demand for spectrum resources has increased dramatically, while the available spectrum resources are very limited. In order to improve spectrum utilization and meet users' needs for high-speed, low-latency communication, researchers have explored spectrum resource allocation schemes based on LAMs and GAI models. \textcolor{black}{For example, Zhang et al. \cite{10679152} proposed a GAI agent-based framework using LLM and RAG to build accurate system models via interactive dialogue. To optimize these models, they introduced an MoE PPO method that combines expert networks with PPO, enabling collaborative decision-making to enhance spectrum efficiency and communication quality.
 In addition, Du et al. \cite{du2024integrated} proposed a GAI and DRL-based framework for optimizing computation offloading and spectrum allocation in 802.11ax Wi-Fi. Combining GDM with the TD3 algorithm and using the Hungarian algorithm for RU allocation, it improved bandwidth utilization, latency, and energy consumption in simulations.}

\subsection{Energy resource optimization}
Energy resource optimization is equally crucial in communication networks, especially in scenarios such as mobile communications and IoT. Traditional energy optimization methods are often based on heuristic rules or simple algorithms, which struggle to achieve optimal results in complex and dynamic environments. Researchers are actively exploring energy resource optimization schemes based on GAI models and LAMs to achieve low energy consumption and high efficiency in wireless communications. \textcolor{black}{For example, Xu et al. \cite{xu2024generative} proposed a GAI-based mobile multimedia network framework for dynamic adaptive streaming, intelligent caching, and energy efficiency optimization, enhancing multimedia content distribution. The framework optimizes resource utilization and reduces energy consumption by considering the value of the GAI model and other indicators. Du et al. \cite{du2023enabling} proposed a wireless edge network framework using AIGC services to optimize energy allocation and improve user experience. By dynamically selecting the optimal ASP with a DRL algorithm, the framework reduces task overload and retransmissions, enhancing energy efficiency and service quality. Simulations showed reduced energy consumption and improved content quality and transmission efficiency.}

\subsection{Summary and lessons learned}
\subsubsection{Summary}
 \textcolor{black}{In this chapter, we summarize the application of LAMs in computing and spectrum resource allocation, and energy resource optimization. LAMs can intelligently allocate resources through real-time prediction and analysis of network demand\textcolor{black}{\cite{10683673}}, and LAMs can also optimize energy usage strategies by learning the energy consumption patterns in communication networks\textcolor{black}{\cite{noh2025adaptive}}.}

 \subsubsection{Lessons learned}
\textcolor{black}{From the chapter, we have learned several important lessons. First, while LAMs can improve optimization efficiency in computational resource allocation, their generalization ability is limited under resource-constrained and dynamic communication environments, potentially resulting in suboptimal or even infeasible allocation strategies\textcolor{black}{\cite{10520929}}. In spectrum resource allocation, although LAMs can assist in enhancing spectrum utilization efficiency, their inference processes often rely on complex combinations of expert networks and scheduling mechanisms, leading to significant computational overhead and difficulty in meeting real-time requirements. Regarding energy resource optimization, LAMs are capable of reducing energy consumption through intelligent caching and flow control. However, their stability and interpretability in generating dynamic scheduling strategies remain insufficient. Therefore, improving the generalization ability of LAMs in resource allocation and optimization, as well as reducing their computational complexity, are key challenges that need to be addressed in future research.}

\section{LAMs for network design and management}
	
LAMs play a vital role in network design and management. Through the powerful ability of generative learning, they can conduct detailed analysis and prediction of network traffic, user behavior, and system performance, and empower existing networks. 

\subsection{Network design} Intelligent network design is the key to ensuring efficient operation of the system and high-quality services.  At present, LAMs are widely used in network design with their powerful generation and data processing capabilities. \textcolor{black}{For example, Huang et al. \cite{huang2024large} proposed an AI-generated network (AIGN) framework using GAI and reinforcement learning for automated network design. It employs a diffusion model to learn design intent and generate customized solutions under multiple constraints, enabling intelligent network design.
Zou et al. \cite{zou2023wireless} proposed a wireless multi-agent GAI network that leverages LLMs on devices for autonomous networking. Multi-agent GAI integration enables network model design, reasoning, multimodal data processing, and resource management.
Huang et al. \cite{huang2024large} proposed ChatNet, a domain-adaptive network LLM framework that uses natural language for intelligent network design, diagnosis, configuration, and security. ChatNet automates tasks by pre-training and fine-tuning open-source LLMs to understand network language and access external tools like simulators, search engines, and solvers.}

\subsection{Network management}

Traditional data processing methods are difficult to meet the 6G network's requirements for massive data, complex tasks, and real-time performance, and the emergence of LAMs provides new ideas for solving these problems.
\textcolor{black}{For example, Wang et al. \cite{wang2023network} proposed NetLM, a network AI architecture using ChatGPT for network management and optimization. Based on LLM, NetLM analyzes network packet sequences and dynamics, unifying network indicators, traffic, and text data through multimodal representation learning to enhance data processing and understand network status, user intent, and complex patterns in 6G networks.
Dandoush et al. \cite{dandoush2024large} proposed a framework combining LLMs with multi-agent systems for network slicing management. Network slicing enables virtual networks on shared infrastructure, but current methods struggle with complex service requirements. The framework uses LLMs to translate user intents into technical requirements and a multi-agent system for cross-domain collaboration, enabling efficient slice creation and management. It also addresses challenges like data acquisition, resource demands, and security.
Yue et al. \cite{yue2023ai} proposed a LAM-enabled 6G network architecture that enhances management efficiency by extracting insights from heterogeneous data. LAMs automate operations, maintenance, and reasoning tasks, reducing human intervention. Using edge computing, LAMs process data in high-concurrency scenarios, improving performance and resource scheduling. The study also addresses challenges like data governance and computational resource needs in 6G networks.
}

\subsection{Summary and lessons learned} 
\subsubsection{Summary}
\textcolor{black}{This chapter summarizes the application of LAMs in network design and management, including the optimization of network architecture design and the management of network slicing. LAMs use their powerful data processing and generation capabilities to efficiently process large amounts of data in 6G networks and realize intelligent network design and management\textcolor{black}{\cite{10700707}\cite{chen2025llm}}.}

\subsubsection{Lessons learned}	

\textcolor{black}{From the chapter, we have learned
several important lessons. First, in network design, although LAMs can automatically generate customized network solutions that satisfy multiple constraints by learning network intents, ensuring the feasibility and stability of the design under multi-constraint conditions remains a significant challenge. Moreover, in network management, LAMs can enhance capabilities in network state analysis, user intent understanding, and data pattern learning, yet difficulties persist in handling large-scale heterogeneous data and meeting real-time performance requirements\textcolor{black}{\cite{huang2024large}}.}

\section{LAMs for edge intelligence}	

LAMs have a wide range of application scenarios for improving edge intelligence. In the following, we discuss the three aspects of LAMs for edge intelligence.

\subsection{Edge training and application of LAMs} 
Edge LAMs are widely used in edge devices due to their ease of deployment while retaining strong data processing capabilities.
\textcolor{black}{For example, the Edge-LLM framework proposed by Yu et al. \cite{yu2024edge} addresses computational and memory overheads in adapting LLMs on edge devices. It uses a layered unified compression (LUC) technique to optimize model compression, reduces memory usage through adaptive layer tuning, and introduces hardware scheduling to handle irregular computation patterns.}

\textcolor{black}{The EdgeShard framework proposed by Zhang et al. \cite{zhang2024edgeshard} distributes LLMs across multiple devices using model sharding. It employs a dynamic programming algorithm to optimize device selection and model partitioning, balancing inference latency and throughput. Experimental results demonstrate a 50\% reduction in latency and a doubling of throughput, providing an efficient solution for LLMs inference in collaborative edge computing.}
\textcolor{black}
{Qu et al. \cite{qu2025mobile} reviewed the integration of LLMs with mobile edge intelligence (MEI) and proposed the MEI4LLM framework to improve deployment efficiency in edge environments. The paper covers technologies like caching, distributed training, and inference, and discusses future directions such as green computing and secure edge AI. It highlights the importance of edge intelligence for low-latency and privacy-sensitive tasks, providing a theoretical foundation for broader LLM applications.}

\textcolor{black}{Xu et al. \cite{xu2023federated} explored federated fine-tuning of LLMs and FM in edge networks, focusing on memory bandwidth limitations. They used energy efficiency to measure computational efficiency, comparing it with model FLOP utilization (MFU). Results show that energy efficiency is better for real-time monitoring, with optimal efficiency achieved on embedded devices using smaller batch sizes.}

\textcolor{black}{In addition, \textcolor{black}{Zhao et al. \cite{zhao2024edge} proposed a framework for LLM deployment that combines edge and terminal collaboration, using serial inference on terminals and parallel inference on edge servers. This reduces latency and optimizes energy consumption, improving model performance across different network conditions and offering an efficient solution for LLM deployment in wireless networks.}
}
Khoshsirat et al. \cite{khoshsirat2024decentralized} studied the application of decentralized LLM inference on energy-constrained edge devices and proposed an inference framework that integrates energy harvesting, enabling distributed devices to collaboratively perform model inference tasks.
\textcolor{black}{Lin et al. \cite{lin2023pushing} explored deploying LLMs in 6G edge environments, addressing computational load with split learning, quantization, and parameter-efficient fine-tuning. The paper proposed tailored LLMs training and inference strategies for edge environments, providing a research path for distributed AI in 6G networks.}
\textcolor{black}{Rong et al. \cite{rong2024large} proposed the LSGLLM-E architecture for large-scale traffic flow prediction, addressing spatiotemporal correlation issues in road networks. The method reduces central cloud pressure by decomposing the network into sub-networks and using RSUs as edge nodes for computation. The LSGLLM model captures dynamic spatiotemporal features, overcoming the limitations of existing LLMs in large-scale road network predictions.}

\subsection{Edge resource scheduling meets LAMs}

\textcolor{black}{Edge devices face limitations in computational power and storage, while LAMs require efficient computation, real-time response, and low-latency data transmission. This creates two main challenges: (1) how to effectively allocate resources to ensure the efficient operation of LAMs on edge devices, and (2) how to leverage the powerful optimization capabilities of LAMs to design improved edge resource scheduling strategies.
To address these, various solutions have been proposed, integrating task offloading, computational, and storage resource optimization to enhance edge device performance in AI tasks.}
\textcolor{black}{For example, Friha et al. \cite{friha2024llm} analyzed LLM-based edge intelligence optimization in resource-constrained environments and proposed strategies to address computing and storage limitations. Techniques like model compression, memory management, and distributed computing enable efficient LLM operation on edge devices. These optimizations improve deployment efficiency and expand LLM applications in areas like personalized medicine and automation.}
\textcolor{black}{Dong et al. \cite{dong2023lambo} proposed the LAMBO framework for LLM-based mobile edge computing (MEC) offloading, addressing challenges in traditional deep offloading architectures, such as heterogeneous constraints and local perception. The framework uses an input embedding (IE) model to convert task data and resource constraints into embeddings, and an asymmetric encoder-decoder (AED) model to extract features and generate offloading decisions and resource allocations.}
\textcolor{black}{Lai et al. \cite{lai2024resource} proposed the GMEN framework to enhance the intelligence and efficiency of mobile edge networks in the 6G era. By combining GAI with edge networks and using methods like model segmentation, the framework offloads AI tasks to reduce network burden. The Stackelberg game model is applied to optimize resource allocation and encourage edge devices to contribute computing resources, reducing overhead.}

 \subsection{Federated learning of LAMs}
	
\textcolor{black}{FL protects privacy and reduces reliance on centralized resources by training models locally, but traditional small models have limited capabilities. The emergence of LAM, with its powerful representation capabilities, enables FL to handle more complex tasks without centralized data, significantly improving personalized services and prediction accuracy. }

\textcolor{black}{For example, Xu et al. \cite{xu2023fwdllm} proposed FwdLLM, a federated learning protocol that enhances LLM on mobile devices using a backpropagation-free training method. FwdLLM combines efficient parameter fine-tuning techniques like LoRA and adapters to distribute the computational load, improving memory and time efficiency, and enabling LLM fine-tuning on ordinary commercial mobile devices.
Peng et al. \cite{peng2024personalized} proposed a personalized semantic communication system using GAI, enhancing performance through personalized local distillation (PLD) and adaptive global pruning (AGP). PLD allows devices to select models based on local resources and distill knowledge into a simpler model for FL. AGP improves the global model by pruning it based on the communication environment, reducing energy consumption and improving efficiency. Through these innovative methods, the application of LAM in personalized FL has demonstrated significant advantages.
In addition, Jiang et al. \cite{jiang2024personalized} proposed two personalized wireless federated fine-tuning methods: personalized federated instruction tuning (PFIT) and personalized federated task tuning (PFTT). PFIT uses reinforcement learning with human feedback for personalization, while PFTT combines a global adapter with LoRA to reduce communication overhead and accelerate fine-tuning, addressing privacy, data heterogeneity, and high communication challenges in wireless networks.}

\subsection{Summary and lessons learned}       
  
\subsubsection{Summary}
\textcolor{black}{This chapter summarizes the application of LAMs in edge intelligence, including the edge training and application of LAMs, resource management and scheduling, as well as federated learning of LAMs. Through edge training and the application of LAMs, the performance of LAMs on edge devices can be effectively enhanced. Resource management and scheduling enable dynamic resource allocation through LAMs\textcolor{black}{\cite{10591707}}. LAMs enables federated learning to tackle more complex tasks without centralized data, improving prediction accuracy and enhancing personalized services\textcolor{black}{\cite{10944018}}.
}

 \subsubsection{Lessons learned}
\textcolor{black}{From the chapter, we have learned
several important lessons. First, when training and deploying LAMs on edge devices, the limited computational and memory resources pose a significant barrier to their widespread adoption\textcolor{black}{\cite{10333824}}. This issue is particularly critical in latency-sensitive applications, where effectively reducing model parameters and optimizing computation patterns remains an urgent area for further investigation. In addition, federated learning of LAMs faces challenges such as limited resources, data heterogeneity, and personalization, with future research focusing on efficient collaboration, robust optimization, and privacy-preserving personalization.}

 \section{LAMs for semantic communication}

 \textcolor{black}{The rapid advancement of communication technology is continuously propelling human society towards higher levels of intelligence. In particular, the emergence of LAMs has profoundly revolutionized the design and optimization of communication systems, shifting the paradigm from traditional data communication to semantic communication. This transformation extends beyond signal transmission to encompass information comprehension, unlocking a wide range of potential application scenarios. We present an overview of the related work of LAMs in semantic communications in the following sections.}

\subsection{LLM-based semantic communication systems}
\textcolor{black}{LLMs, with their powerful natural language understanding and generation capabilities, can perform semantic-level analysis and processing in complex communication environments, significantly enhancing the intelligence of semantic communication systems. Especially in future networks like 6G, LLMs can support more efficient and flexible semantic communication architectures, enabling intelligent applications of semantic communication.}
\textcolor{black}{For example, Wang et al. \cite{wang2024large} proposed a semantic communication system framework based on LLMs, applying them directly to physical layer encoding and decoding. The system leveraged LLM training and unsupervised pre-training to build a semantic knowledge base, used beam search algorithms to optimize decoding and reduce complexity, and required no additional retraining or fine-tuning of existing LLMs.}
Jiang et al.\cite{jiang2024large6} proposed a large generative model-assisted talking face semantic communication system (LGM-TSC) to address the challenges in talking face video communication, including low bandwidth utilization, semantic ambiguity, and degraded quality of experience (QoE). The system introduces a generative semantic extractor (GSE) based on the FunASR model at the transmitter, which converts semantically sparse talking face video into high information density text. A private knowledge base (KB) based on LLMs is used for semantic disambiguation and correction, complemented by a joint knowledge base semantic channel coding scheme. At the receiver side, the generative semantic reconstructor (GSR) using BERTVITS2 and SadTalker models converts the text back into a high-QoE talking face video that matches the user’s voice tone.
Chen et al.\cite{chen2024semantic} proposed a novel semantic communication framework based on LLMs to address the challenges in underwater communication, including semantic information mismatch and the difficulty of accurately identifying and transmitting critical information. The framework leverages visual LLMs to perform semantic compression and prioritization of underwater image data, selectively transmitting high-priority information while applying higher compression rates to less important areas. At the receiver side, the LLMs-based recovery mechanism works in conjunction with global visual control networks and key region control networks to reconstruct the image, improving communication efficiency and robustness. The system reduces the overall data size to 0.8\% of the original data.

\textcolor{black}{In addition, Jiang et al.\cite{jiang2024semantic} proposed a method for integrating FMs, including LLMs, across the validity, semantic, and physical layers in semantic communication systems. This integration leverages general knowledge to alter system design, thereby improving semantic extraction and reconstruction. The study also explored the use of compact models to balance performance and complexity, and compared three approaches using FMs. The research emphasizes the need for further analysis of the impact of FMs on computational and memory complexity, as well as the unresolved issues that require attention in this field.
Kalita et al.\cite{kalita2024large} proposed a framework that integrates LLMs with semantic communication at the network edge to enable efficient communication in IoT networks. The framework leverages the capabilities of LLMs, training on diverse datasets with billions of parameters, to improve communication performance in scenarios where current technologies are approaching the Shannon limit. The system is designed to run on near-source computing technologies such as edge, thereby enhancing communication efficiency in IoT environments.
Wang et al.\cite{10570717} proposed a general end-to-end learning semantic communication model using LLMs to enhance the performance of next-generation communication systems. The model combines subword-level tokenization, a gradient-based rate adapter to match the rate requirements of any channel encoder/decoder, and fine-tuning to incorporate private background knowledge.}

\subsection{Other LAM-based semantic communication systems}
\textcolor{black}{In addition to LLMs, research on semantic communication systems based on other LAMs also plays a crucial role in advancing the intelligence of semantic communication systems\cite{jiang2025lightweight}\cite{jiang2025m4sc}.}
For example, Jiang et al.\cite{jiang2024visual} proposed a novel cross-modal semantic communication system based on VLM (VLM-CSC) to address the challenges in image semantic communication, such as low semantic density in dynamic environments, catastrophic forgetting, and uncertain signal-to-noise ratio. The VLM-CSC system includes three key components: (1) a cross-modal knowledge base, which extracts high-density textual semantics from semantically sparse images at the transmitter side and reconstructs the original image at the receiver side to alleviate bandwidth pressure; (2) an memory-augmented encoder and decoder, which employs a hybrid long/short-term memory mechanism to prevent catastrophic forgetting in dynamic environments; and (3) a noise attention module, which adjusts semantic and channel coding based on SNR to ensure robustness. \textcolor{black}{ Zhang et al. \cite{zhang2024addressing} proposed the "Plan A - Plan B" framework using MLLMs to address the out-of-distribution (OOD) problem in image semantic communication. It leverages MLLMs' generalization to assist traditional models during semantic encoding. A Bayesian optimization scheme reshapes MLLM distributions by filtering irrelevant vocabulary and using contextual similarity as prior knowledge. At the receiver, a "generate-critic" framework improves reconstruction reliability, addressing the OOD problem and enhancing semantic compression.
Jiang et al. \cite{jiang2024large1} proposed the GAM-3DSC system to address challenges in 3D semantic extraction, redundancy, and uncertain channel estimation in 3D scene communication. By introducing LVM, the system enables user-driven 3D semantic extraction, adaptive multi-view image compression, and CSI estimation and optimization for effective target-oriented 3D scene transmission.
Xie et al. \cite{xie2024towards} proposed a new semantic communication architecture that integrates large models by introducing a memory module. This enhances semantic and contextual understanding, improves transmission efficiency, and addresses spectrum scarcity.}

\textcolor{black}{Yang et al. \cite{yang2024rethinking} proposed the "M2GSC" framework for generative semantic communication in multi-user 6G systems. It employs MLLMs as a shared knowledge base (SKB) for task decomposition, semantic representation standardization, and translation, enabling standardized encoding and personalized decoding. The framework also explores upgrading the SKB to a closed-loop agent, adaptive encoding offloading, and multi-user resource management.
Do et al. \cite{do2024lightweight} proposed a mamba-based multi-user multimodal deep learning semantic communication system to enhance efficiency in resource-constrained networks. By replacing the transformer with mamba architecture, the system improves performance and reduces latency. It introduces a new semantic similarity metric and a two-stage training algorithm to optimize bit-based metrics and semantic similarity.
}
Jiang et al.\cite{jiang2024large5} proposed a multimodal semantic communication (LAM-MSC) framework based on LAMs to address the challenges in multimodal semantic communication, such as data heterogeneity, semantic ambiguity, and signal distortion during transmission. The framework includes multimodal alignment (MMA) based on MLM, which facilitates the conversion between multimodal and unimodal data while maintaining semantic consistency. It also introduces a personalized knowledge base (PKB) based on LLMs to perform personalized semantic extraction and recovery, thereby resolving semantic ambiguity. Additionally, a channel estimation method based on conditional GANs is used to estimate wireless CSI, mitigating the impact of fading channels on semantic communication.

\subsection{Summary and lessons learned}

\subsubsection{Summary}{
\textcolor{black}{This chapter summarizes the applications of LAMs in semantic communications, including LLMs and other LAMs. The powerful data processing capabilities of LAMs can effectively reduce communication overhead\textcolor{black}{\cite{10827255}}, improve communication efficiency, enhance the expression and understanding of semantic information, and enable more flexible, intelligent, and efficient semantic communications\textcolor{black}{\cite{10570717}}. }}

\subsubsection{Lessons learned}

\textcolor{black}{From the chapter, we have learned
several important lessons. First, although LAMs show remarkable performance in semantic extraction and reconstruction when directly applied to physical-layer encoding and decoding, their high computational complexity remains a major bottleneck for real-time deployment in resource-constrained environments\textcolor{black}{\cite{10615340}}. Second, current semantic communication systems have yet to fully address key issues such as semantic information alignment, ambiguity resolution, and bandwidth utilization optimization under dynamic network conditions. This is especially evident in multi-user and multimodal scenarios, where effective semantic standardization and cross-modal collaboration remain open research problems.}

\section{LAM-based agentic systems}
\textcolor{black}{The application of intelligent agentic systems based on LLMs and other GAI models is an important way to address the challenges faced by current communication systems. These intelligent agent-driven systems can improve the transmission efficiency of semantic communication systems and optimize resource allocation of edge devices.} 

\subsection{Agentic systems based on LLMs}
Agentic systems based on LLMs are widely applied in communication systems due to their powerful NLP capabilities. \textcolor{black}{For example, Xu et al. \cite{xu2024large2} proposed a 6G-based LLM agent split learning system to improve the efficiency of local LLM deployment on resource-limited mobile devices. The system enables mobile-edge collaboration through modules for perception, semantic alignment, and context binding. A model caching algorithm enhances model utilization and reduces network costs for collaborative LLM agents.
Jiang et al. \cite{jiang2024large3} proposed a multi-agent system to address challenges LLMs face in 6G communication evaluation, including lack of native data, limited reasoning, and evaluation difficulties. The system comprises multi-agent data retrieval (MDR), cooperative planning (MCP), and evaluation and reflection (MER). A semantic communication system case study demonstrated its effectiveness.
Tong et al. \cite{tong2024wirelessagent} proposed WirelessAgent, which uses LLMs to build AI agents addressing scalability and complexity in wireless networks. With advanced reasoning, multimodal data processing, and autonomous decision-making, it enhances network performance. Applied to network slicing management, WirelessAgent accurately understands user intentions, allocates resources effectively, and maintains optimal performance.
}

\textcolor{black}{In addition, Zou et al. \cite{zou2023wireless} proposed wireless multi-agent GAI networks to overcome cloud-based LLM limitations by enabling task planning through multi-agent LLMs. Their method explores game-theory-based multi-agent LLMs and designs an architecture for these systems. A case study demonstrates how device-based LLMs collaborate to solve network solutions.
Wang et al. \cite{wang2024generative} proposed GAI Agents, a next-generation MIMO design method addressing challenges in performance analysis, signal processing, and resource allocation. By combining GAI agents with LLMs and RAG, the method customizes solutions. The paper discusses the framework and demonstrates its effectiveness through two case studies, improving MIMO system design.
Zhang et al. \cite{10679152} proposed GAI agents for satellite communication network design, tackling system modeling and large-scale transmission challenges. The method uses LLMs and RAG to build interactive models and MoE for transmission strategies. It combines expert knowledge and employs MoE-PPO for simulation, validating GAI agents and MoE-PPO in customized problems.
Wang et al. \cite{wang2024large2} proposed an LLM-powered base station siting (BSS) optimization framework, overcoming limitations of traditional methods. By optimizing prompts and using automated agent technology, the framework improves efficiency, reduces costs, and minimizes manual effort. Experiments show that LLMs and agents enhance BSS optimization.
}

\subsection{Agentic systems based on other GAI models}
In addition to LLMs, agent systems based on other GAI models are also widely applied in the research of communication systems.
\textcolor{black}{For example, Yang et al. \cite{10815060} proposed an agent-driven generative semantic communication (A-GSC) framework based on reinforcement learning to address challenges in remote monitoring for intelligent transportation systems and digital twins in 6G. Unlike previous research on semantic extraction, A-GSC integrates the source information's intrinsic properties with task context and introduces GAI for independent design of semantic encoders and decoders.
Chen et al. \cite{chen2024enabling} proposed a system architecture for AI agents in 6G networks, tackling challenges in network automation, mobile agents, robotics, autonomous systems, and wearable AI agents. This architecture enables deep integration of AI agents within 6G networks and collaboration with application agents. A prototype validated their capabilities, highlighting three key challenges: energy efficiency, security, and AI agent-customized communication, laying the foundation for AI agents in 6G.}

\subsection{Summary and lessons learned} 
\subsubsection{Summary}
\textcolor{black}{This chapter summarizes the research and applications of intelligent agentic systems based on LLMs and other GAI models in communication\textcolor{black}{\cite{10839354}}. By leveraging the powerful data analysis and processing capabilities of these technologies, agentic systems can more effectively address the challenges faced by the current communication system, thereby enabling more efficient information transmission\textcolor{black}{\cite{10620727}}.}

\subsubsection{Lessons learned}

\textcolor{black}{From the chapter, we have learned
several important lessons. First, constrained by the computational capabilities of mobile terminals, LAM-based agentic systems face challenges such as low computational efficiency and complex model scheduling in local deployment and collaborative execution. Although some studies have introduced model caching and task partitioning mechanisms to improve resource utilization, the overall system still struggles to meet the demands of high concurrency and low latency in modern communication scenarios\textcolor{black}{\cite{10720863}}. Second, while multi-agent systems can collaboratively accomplish complex tasks—such as data retrieval, planning, and reflection—the lack of domain-specific knowledge and high-quality communication data limits their reasoning and decision-making performance in advanced tasks such as those in 6G semantic communications.
}

\section{LAMs for emerging applications}
\label{sec:emerging}
	The combination of LAMs with the emerging applications is the driving technological innovation in multiple industries and fields. These LAMs use their large data sets and deep learning capabilities to provide strong support for applications such as smart healthcare, carbon emissions, digital twin, artificial intelligence of things (AIoT), integrated satellite, aerial, and terrestrial networks (ISATN), and integration of UAVs and LLMs. In the following, we introduce LAMs for these emerging applications in detail.
	
 \subsection{Smart healthcare}
	
	Smart healthcare uses these advanced technologies to improve the efficiency and quality of medical services. Through data-driven decision support systems, medical institutions can achieve accurate diagnosis and personalized treatment, thereby meeting the needs of patients more effectively. In the smart healthcare, through LAMs and combined with digital twin technology, we always pay attention to the patient's physical condition and provide personalized medical care for patients. The openCHA proposed by Abbasian et al. \cite{abbasian2023conversational} provided users with personalized services in medical consultation. openCHA is an open source framework based on LLM, which aims to provide users with personalized smart healthcare services. The openCHA framework overcomes the limitations of the existing LLM in healthcare, including lack of personalization, multimodal data processing, and real-time knowledge updating, by integrating external data sources, knowledge bases, and AI analysis models.

 \subsection{Carbon emissions}
\textcolor{black}{In controlling carbon emissions, Wen et al. \cite{wen2024generative} proposed a GAI-based low-carbon AIoT solution to reduce carbon emissions from energy consumption in communication networks and computation-intensive tasks. GAI, using GANs, RAG, and GDMs, optimizes resource allocation, reduces energy waste, and improves efficiency. The paper explores GAI applications in energy internet (EI), data center networks, and mobile edge networks. In EI, GAI optimizes renewable energy use; in data centers, it improves the management of information and communication technology (ICT) equipment and cooling systems; and in mobile edge networks, GAI, combined with IRS deployment and semantic communication technologies, reduces power consumption. The findings show GAI’s superiority in carbon emission optimization, supporting low-carbon AIoT and sustainability goals.
}

\subsection{Digital twins} The application of LAMs in digital twins is a key force in promoting the development of this technology. \textcolor{black}{For example, Xia et al. \cite{xia2023towards} proposed a framework integrating LLMs, digital twins, and industrial automation systems for intelligent planning and control of production processes. LLM-based agents interpret descriptive information in the digital twin and control the physical system via service interfaces. These agents function as intelligent agents across all levels of the automation system, enabling autonomous planning and control of flexible production processes.
Hong et al. \cite{hong2024llm} proposed an LLM-based digital twin network (DTN) framework, LLM-Twin, to improve communication and multimodal data processing in DTNs. They introduced a digital twin semantic network (DTSN) for efficient communication and computation and a small-to-giant model collaboration scheme for efficient LLM deployment and multimodal data processing. A native security strategy was also designed to maintain security without sacrificing efficiency. Numerical experiments and case studies validated LLM-Twin’s feasibility.
}

\subsection{Artificial intelligence of things}
\textcolor{black}{In AIoT, Cui et al. \cite{cui2024llmind} proposed the LLMind framework, showing how combining LLMs with domain-specific AI modules enhances IoT device intelligence and collaboration. It automates tasks and enables cooperation through high-level language instructions. A key feature is the language-to-code mechanism that converts natural language into finite state machine (FSM) representations for device control scripts, optimizing task execution. With an experience accumulation mechanism, LLMind improves responsiveness and supports efficient collaboration in dynamic environments, highlighting its potential in IoT intelligent control.
}

\subsection{Integrated satellite, aerial, and terrestrial networks}
\textcolor{black}{Javaid et al. \cite{javaid2024leveraging} explored the potential of incorporating LLMs into ISATNs. ISATNs combine various communication technologies to achieve seamless cross-platform coverage. The study demonstrates that LLMs, with their advanced AI and machine learning capabilities, can play a crucial role in data stream optimization, signal processing, and network management, particularly in 5G/6G networks. The research not only provided a comprehensive analysis of ISATN architecture and components but also discussed in details how LLMs can address bottlenecks in traditional data transmission and processing. Additionally, the paper focused on challenges related to resource allocation, traffic routing, and security management within ISATN network management, highlighting technical difficulties in data integration, scalability, and latency. The study concludes with a series of future research directions aiming at further exploring LLM applications to enhance network reliability and performance, thus advancing the development of global intelligent networks.}

\subsection{Integration of UAVs and LLMs}
\textcolor{black}{Regarding the integration of UAVs and LLMs, Javaid et al. \cite{javaid2024large} conducted a systematic analysis of the current state and future directions of combining LLMs with UAVs. The study thoroughly examined the role of LLMs in enhancing UAV autonomy and communication capabilities, particularly in key areas such as spectrum sensing, data processing, and decision-making. By integrating LLMs, UAVs can achieve a higher level of intelligence in complex tasks, including autonomous responses and real-time data processing. The authors evaluated the existing LLM architectures, focusing on their contributions to improving UAV autonomous decision-making, especially in scenarios like disaster response and emergency communication restoration. Additionally, the paper highlighted the technical challenges faced in future research, emphasizing the importance of further exploring legal, regulatory, and ethical issues to ensure the effective and sustainable integration of LLM and UAV technologies.}

\subsection{Summary and lessons learned}
	
\subsubsection{Summary}
\textcolor{black}{This chapter highlights the role of LAMs in emerging applications. In smart healthcare, LAMs enable personalized care and efficient diagnostics through frameworks like openCHA. For carbon emissions, LAM-enabled optimization frameworks address environmental challenges, holding significant value in achieving sustainability and carbon
neutrality goals.
\textcolor{black}{In digital twins, LAMs significantly advance their development in industrial automation and other domains by enhancing intelligent perception, communication\textcolor{black}{\cite{10757470}}, and control capabilities.} In AIoT, LAMs enhance device collaboration, task execution, and user interaction\textcolor{black}{\cite{10729865}}. Additionally, LAMs contribute to networking technologies like ISATNs and UAVs by improving resource allocation, decision-making, and communication. These applications illustrate LAMs' growing influence in addressing complex challenges across various fields.}
\subsubsection{Lessons learned}

\textcolor{black}{From the chapter, we have learned
several important lessons. One primary issue is the insufficiency of data quality and diversity, which hampers the generalization ability of LAMs across different domains. For instance, in smart healthcare, while LAMs can enhance the accuracy of personalized medicine, data privacy constraints often limit data sharing, potentially introducing biases into the model. In carbon emission optimization and AIoT scenarios, LAMs rely heavily on high-quality real-time data, and issues such as data incompleteness or latency can negatively affect optimization outcomes. Furthermore, security and privacy concerns are critical. In digital twin applications, decisions made by LAMs can directly impact the operation of physical systems, and any data tampering or model attacks could lead to severe consequences. This risk is particularly pronounced in integrated applications involving ISATNs and UAVs with LAMs\textcolor{black}{\cite{10839306}}, where cybersecurity vulnerabilities could be exploited maliciously, resulting in data breaches or communication disruptions.}

\section{Research challenges}
\label{sec:challenges and future}

\textcolor{black}{Although the LAMs have great application potential for communications, they still face many challenges. This section mainly introduces some research challenges and potential solutions of LAMs in communication.}
\subsubsection{Lack of high-quality communication data}
\textcolor{black}{In the application of cutting-edge technologies such as 6G and the IoE, data acquisition and diversity pose significant challenges. This issue is particularly critical in core tasks such as wireless communication, interference mitigation, and spectrum management, where the lack of high-quality labeled data constrains the training efficacy of LAMs. First, the cost of data collection is high, especially in complex network environments where substantial investments in hardware and sensors are required, increasing both equipment expenditures and long-term maintenance complexity. Second, data privacy and ethical concerns have become increasingly prominent, with stringent privacy regulations imposing strict constraints on data collection, thereby complicating the acquisition of effective datasets. Lastly, the scarcity of labeled data presents a major limitation, particularly for high-precision tasks, as obtaining labeled data requires domain expertise and expensive equipment. Moreover, the dynamic nature of communication environments makes it difficult to comprehensively cover all conditions, ultimately restricting the generalization capability of models. The data scarcity issue in communication hinders the application of LAMs. To address this challenge, techniques such as data augmentation, self-supervised learning, and GANs can be employed to expand dataset size, improve training efficiency, and reduce reliance on high-quality labeled data. These approaches enable LAMs to better adapt to dynamic communication scenarios.}  

\subsubsection{Lack of structured communication knowledge}
\textcolor{black}{LAMs struggle to solve complex communication problems due to their limited understanding of communication theory, protocols, and standards. As LAMs primarily rely on data-driven learning, their decision-making is often based solely on statistical patterns extracted from training data, neglecting the structured knowledge inherent in communication. For instance, factors such as signal attenuation, interference, and noise directly impact communication system design. However, LAMs find it challenging to embed these complex structured knowledge elements, particularly in tasks such as interference cancellation, spectrum allocation, and channel modeling. This limitation often results in an inability to accurately capture physical constraints, ultimately affecting overall system performance. To overcome the challenge of lacking structured communication knowledge, researchers can integrate LAMs with communication principles through physics-informed networks and construct structured communication knowledge using knowledge graphs. By combining domain-specific expertise with the reasoning capabilities of LAMs, these approaches enhance model performance in complex communication scenarios.}

\subsubsection{Generative hallucination in communication}
\textcolor{black}{Hallucination in LAMs has emerged as a significant challenge in communication. This phenomenon can be categorized into two major types: factual hallucination, where the model generates incorrect content that deviates from the correct results, and faithfulness hallucination, where the model fails to follow user instructions accurately, producing irrelevant or inconsistent responses. The root cause of these hallucinations lies in the model's data-driven training process, which lacks a deep understanding of communication system principles. As a result, inaccurate decisions may arise in tasks such as signal quality prediction and network optimization, severely degrading network performance and user experience. To address this issue, several strategies can be employed to enhance the accuracy and stability of model outputs. These include incorporating the physical constraints of communication systems, leveraging traditional optimization methods to assist model outputs, employing ensemble decision-making across multiple models to improve output consistency, and designing specialized hallucination detection and mitigation algorithms. By ensuring that the outputs align with the objective principles of communication systems, these approaches enhance the LAM’s reliability and applicability in real-world communication scenarios.}

\subsubsection{Limitations of reasoning ability}
\textcolor{black}{LAMs in communication systems primarily rely on data-driven pattern recognition and prediction. However, when faced with communication tasks requiring high levels of abstraction and multi-step reasoning, they often struggle to accurately comprehend complex logical relationships, leading to unreliable decision-making. In scenarios such as wireless channel modeling, spectrum allocation, and interference management, LAMs must infer multiple interdependent physical parameters and network factors to make well-informed decisions. Without deep reasoning capabilities, LAMs may fail to properly account for these intricate dependencies, resulting in outputs that contradict the physical principles governing real-world communication systems. To address the reasoning limitations of LAMs in handling complex communication problems, techniques such as tree-of-thought reasoning, graph-based reasoning, and long-chain reasoning can be employed. These approaches leverage hierarchical structured information, multi-step inference, and process-level reward functions to enhance logical reasoning, improve decision accuracy, and increase model adaptability. By integrating these advanced reasoning mechanisms, LAMs can become more efficient and precise in tackling complex communication tasks.}

\subsubsection{Poor explainability in LAMs}
\textcolor{black}{The black-box nature of LAMs in communication presents a critical challenge due to their poor explainability. The internal mechanisms and decision-making processes of these models are often opaque, making it difficult to trace decisions in tasks such as fault diagnosis, system optimization, and network management, thereby increasing the complexity of troubleshooting. Additionally, the lack of explainability raises ethical and legal concerns, particularly in areas involving user privacy and network security. To address this issue, explainable AI (XAI) techniques can be employed to enhance the transparency and trustworthiness of LAMs. Methods such as local interpretable model-agnostic explanations (LIME) and shapley additive explanations (SHAP) can help users understand the rationale behind model decisions. Furthermore, visualizing the model’s decision-making process through graphical representations can provide insights into reasoning pathways. These solutions not only improve explainability but also enable a transparent and traceable decision-making process for communication systems, enhancing both trust and operational reliability.}
\subsubsection{Adaptability in dynamic environments}
\textcolor{black}{Due to the dynamic variations in network topology, channel conditions, and user demands, communication systems face significant challenges in optimization and management, making rapid adaptation and real-time decision-making crucial. While LAMs demonstrate strong performance in static environments, their adaptability in dynamic scenarios often becomes a bottleneck in practical applications. In tasks such as wireless channel estimation, resource scheduling, and interference cancellation, LAMs must swiftly respond to environmental changes to ensure accurate and timely predictions. If the model fails to adjust its generative capabilities in response to evolving network conditions and user requirements, it may lead to delayed or inaccurate predictions, thereby degrading system performance. To address this issue, techniques such as online learning, continual learning, multi-task learning, and meta-learning offer effective solutions. These approaches enable LAMs to dynamically optimize parameters, adapt in real-time, and leverage knowledge transfer across tasks, thereby enhancing their reasoning ability, adaptability, and robustness in dynamic communication environments.}

\subsubsection{Diversity of communication tasks}
\textcolor{black}{The field of communications encompasses a wide range of highly specialized tasks, including signal processing, network optimization, interference mitigation, and spectrum management. These tasks differ significantly in terms of objectives, constraints, and optimization strategies, and are often intricately interrelated. Although LAMs exhibit strengths in multi-task learning, their lack of domain-specific knowledge, variations in optimization requirements, and inconsistencies across tasks make it challenging to adapt to the diverse nature of communication tasks. For instance, signal processing demands a deep understanding of modulation and demodulation techniques, while network optimization focuses on bandwidth allocation and traffic control. Thus, designing model architectures that can flexibly accommodate different communication tasks remains a major challenge. Approaches such as task-specific models, MoE, and transfer learning have shown promise in enhancing the performance of LAMs in this context. Task-specific models allocate dedicated sub-models to different tasks to minimize interference and improve effectiveness; MoE dynamically selects expert models tailored to specific tasks, boosting multi-task learning efficiency; and transfer learning facilitates knowledge transfer, improving the adaptability and generalization of LAMs. These methods collectively enhance the adaptability, efficiency, and accuracy of LAMs in multi-task environments, thereby strengthening their performance and reliability across diverse telecommunication tasks.}
\subsubsection{Resource constraints at the edge}
\textcolor{black}{In mobile devices, edge computing platforms, and IoT devices, hardware resources are typically limited and cannot meet the high computational and energy demands of LAMs. These devices—especially nodes and terminals at the edge of 6G networks—are expected to operate under low-power and resource-constrained conditions, yet their processing capabilities, memory, and energy efficiency fall short of what LAMs require. Direct deployment of LAMs at the edge often results in performance degradation, increased latency, and compromised communication quality and user experience. To improve the efficiency of LAMs on devices with limited computation, storage, and power, several strategies can be employed: model distillation transfers knowledge from LAMs to smaller ones to enhance adaptability; model compression techniques such as pruning and quantization reduce computational and memory overhead; and hardware acceleration leverages specialized hardware like GPUs, TPUs, and FPGAs to speed up inference while lowering power consumption. These approaches effectively enhance the inference efficiency and performance of LAMs in edge and IoT scenarios.}
\subsubsection{High inference latency}
\textcolor{black}{In wireless communications, low latency and high throughput are critical, especially for real-time applications such as autonomous driving and remote healthcare. However, due to their large-scale architectures and complex computational demands, LAMs often suffer from high inference latency, which can lead to delayed system responses, reduced throughput, instability in mission-critical tasks, and inefficient resource utilization. As communication systems grow increasingly complex, a key challenge is to reduce inference latency while maintaining model accuracy. To address the issue of high inference latency, several optimization techniques can be applied. Operator fusion reduces memory access and data transfer delays by combining multiple operations, thereby improving computational efficiency. Speculative sampling accelerates inference by predicting future steps in advance, reducing computational overhead. These methods effectively lower latency, enhance response time, and improve resource utilization, ensuring that LAMs can meet the stringent performance requirements of next-generation communication systems.}

\subsubsection{Security and privacy}
\textcolor{black}{In 6G networks, the use of LAMs for data processing introduces significant security and privacy risks. Since LAMs are often pre-trained in a centralized manner, they are highly susceptible to data breaches, potentially allowing attackers to reconstruct sensitive information. Additionally, data transmissions are vulnerable to man-in-the-middle attacks, eavesdropping, and tampering. Moreover, LAMs themselves may be exposed to adversarial attacks, leading to incorrect predictions and decisions that can compromise network stability. With increasingly stringent data privacy regulations, LAMs must comply with privacy protection requirements to mitigate legal risks and maintain user trust. To address these challenges, researchers have proposed several solutions. Federated learning enables model training on local devices, minimizing the transmission and storage of sensitive data, thereby reducing data exposure risks. Encrypted computing techniques, such as homomorphic encryption and secure multi-party computation, ensure data security even in untrusted environments. These approaches help mitigate security threats associated with large models, enhancing model reliability and user trust, thereby fostering the deep integration of LAMs with next-generation communication technologies.}

\section{Conclusion}
\textcolor{black}{This paper provides a comprehensive review of the development, key technologies, application scenarios, and research challenges of LAMs in communication. It systematically summarizes the critical roles and potential of LAMs, ranging from fundamental theories to practical applications, particularly in the era of 6G, when the demand for efficient, stable, and intelligent communication systems is growing.
First, the paper delves into the foundational aspects of LAMs, including model architectures, classification of different types of LAMs, training paradigms, evaluation methodologies, and optimization mechanisms in communication.
Second, it presents a detailed overview of recent research progress on applying LAMs across various scenarios. The paper systematically analyzes the adaptability and technical advantages of different LAMs in diverse application scenarios, supported by extensive case studies and discussions on cutting-edge developments.
Finally, this paper conducts an in-depth analysis of the key challenges currently faced by LAMs in the communication domain. These challenges include the lack of high-quality communication data, the absence of structured domain knowledge, and the occurrence of generative hallucinations during communication tasks. In addition, limitations such as inadequate reasoning capabilities, poor interpretability, weak adaptability to dynamic environments, and the increased modeling complexity introduced by task diversity further hinder the development of LAMs in communication. Practical deployment is also constrained by limited computational resources at the edge, high inference latency, and critical issues related to data security and privacy protection. It further proposes potential solutions to address these challenges.
Through these efforts, LAMs are expected to enable more intelligent, efficient, and secure services, thereby driving the advancement of 6G and future communication networks.
}

\bibliographystyle{IEEEtran}
\bibliography{ref}

	\newpage
\end{document}